\documentclass[aps,twocolumn,preprintnumbers,amsmath,amssymb,floatfix,superscriptaddress]{revtex4}
\usepackage{graphicx}
\usepackage{color}
\def\tHp{\tau_{\rm Hp}}

\def\SHp{S_{\rm Hp}}
\def\Smax{S_{\rm max}}
\def\RHp{Re_{\rm Hp}}

\def\qRes{q_{\rm s}}
\def\qMin{q_{\rm min}}
\def\rs{r_{\rm s}}
\def\rsa{r_{\rm s1}}
\def\rsb{r_{\rm s2}}

\def\flPsi{\widetilde{\psi}}
\def\flPhi{\widetilde{\phi}}
\def\flVor{\widetilde{u}}

\def\eqPsi{\overline{\psi}}
\def\eqPhi{\overline{\phi}}
\def\eqVor{\overline{u}}
\def\eqCur{\overline{j}}

\def\laplR{\frac{1}{r}\partial_r r\partial_r}

\def\gLin{\gamma_{\rm lin}}

\def\gPeak{\gamma_{\rm peak}}
\def\mPeak{m_{\rm peak}}

\def\Ma{M^{(1)}}
\def\Mb{M^{(2)}}

\definecolor{gray}{rgb}{0.5,0.5,0.5}%
\definecolor{dred}{rgb}{0.5,0.0,0.0}%
\definecolor{dgreen}{rgb}{0.0,0.5,0.0}%

\begin{document}

\preprint{physics/0503068}

\title{Fast growing double tearing modes in a tokamak plasma}

\author{Andreas~Bierwage}
\email{bierwage@center.iae.kyoto-u.ac.jp}
\affiliation{Graduate School of Energy Science, Kyoto University,
Gokasho, Uji, Kyoto 611-0011, Japan}
\author{Sadruddin~Benkadda}
\email{benkadda@up.univ-mrs.fr}
\affiliation{Equipe Dynamique des Syst\`{e}mes Complexes, UMR 6633
CNRS-Universit\'{e} de Provence, 13397 Marseille, France}
\author{Satoshi~Hamaguchi}
\email{hamaguch@ppl.eng.osaka-u.ac.jp}
\affiliation{Graduate School of Engineering, Osaka University, 2-1
Yamadaoka, Suita, Osaka 565-0871, Japan}
\author{Masahiro~Wakatani}
\thanks{deceased}
\affiliation{Graduate School of Energy Science, Kyoto University,
Gokasho, Uji, Kyoto 611-0011, Japan}

\date{\today}

\begin{abstract}
Configurations with nearby multiple resonant surfaces have broad
spectra of linearly unstable coupled tearing modes with dominant high
poloidal mode numbers $m$. This was recently shown for the case of
multiple $q = 1$ resonances [Bierwage \textit{et al.},
Phys. Rev. Lett. \textbf{94} (6), 65001 (2005)]. In the present work,
similar behavior is found for double tearing modes (DTM) on resonant
surfaces with $q \geq 1$. A detailed analysis of linear instability
characteristics of DTMs with various mode numbers $m$ is performed
using numerical simulations. The  mode structures and dispersion
relations for linearly unstable modes are calculated. Comparisons
between low- and higher-$m$ modes are carried out, and the roles of
the inter-resonance distance and of the magnetic Reynolds number
$\SHp$ are investigated. High-$m$ modes are found to be destabilized
when the distance between the resonant surfaces is small. They
dominate over low-$m$ modes in a wide range of $\SHp$, including
regimes relevant for tokamak operation. These results may be readily
applied to configurations with more than two resonant surfaces.
\end{abstract}

\maketitle

\thispagestyle{empty}

%===========================================
\section{Introduction}

Double tearing modes (DTM) are coupled tearing modes on adjacent
resonant surfaces which effectively ``drive each other''
\cite{White77, Pritchett80, Mahajan82}. The resulting instability is
stronger than a single tearing mode (STM), whereby modes with high
poloidal mode numbers $m$ may become dominant \cite{Bierwage05a}.

DTMs are thought to be involved in a variety of dynamical processes in
tokamak plasmas. In the current-ramp-up phase of the tokamak operation
they were used to explain the anomalously strong current penetration
\cite{Schmidt71, Furth73, Stix76} and short-wavelength MHD activity
(e.g., Mirnov oscillations) \cite{Furth73, Goodall84}. Experimental
observations of partial and compound sawtooth crashes (internal
disruptions) \cite{Edwards86, Taylor86, Campbell86, Kim86, Ishida88}
may be explained through kink-tearing modes on double/multiple
resonant surfaces \cite{Carreras79, Parail80, Pfeiffer85,
Chang96}. Phenomena associated with low-beta disruptions in the
presence of double resonant surfaces with $\qRes > 1$ may also
be understood in terms of DTM activity \cite{Carreras79, Persson94,
Ishii02}. In some way related to DTMs is the coupling
between modes on resonant surfaces with different $\qRes$, which
may explain some salient features of major disruptions \cite{Stix76,
Waddell79, Connor88}. A more recent application is related to the
formation of internal transport barriers (ITB) in advanced
tokamaks. DTMs were suggested to be involved in this process due to
the fact that ITBs are often observed in the vicinity of resonant
surfaces or near $\qMin$ \cite{Connor04, Guenter00, Voitsekhovitch02}.

It was found that DTMs with any mode number $m$ become similar to
$m=1$ internal kink modes \cite{Coppi76} when the distance between the
resonant surfaces is sufficiently small \cite{Furth73}. This is
reflected by the linear growth rate $\gLin$ scaling with the magnetic
Reynolds number $\SHp$ as $\gLin \propto \SHp^{-1/3}$
\cite{Pritchett80}. When the distance between the resonant surfaces is
large, the structure of the instability resembles that of individual
STMs localized on each resonant surface. In this case, the scaling law
is $\gLin \propto \SHp^{-3/5}$ \cite{Furth63}. Furthermore, it was
found that the coupling in a DTM may also destabilize linearly stable
tearing modes \cite{Persson94} and speed up the growth of magnetic
islands in the nonlinear regime \cite{White77, Yu96, Ishii00,
Ishii02}. The instability of DTMs may be enhanced by effects like
anomalous electron viscosity \cite{Furth73, Dong03}, finite beta and
sheared toroidal flows \cite{Held99}. Poloidal shear flows, on the
other hand, may have a stabilizing effect \cite{Ofman92, Persson94,
Shen98}, raising the possibility of a dynamical interplay between
shear flows and DTMs \cite{Persson94, Guenter99}. This idea is
supported by recent studies which indicate that MHD activity
associated with multiple resonant surfaces may produce a significant
amount of sheared poloidal flows \cite{Held99, Dong03}.

Until now, studies related to DTMs in cylindrical (or toroidal)
plasmas were focussing mainly on the role of the modes with the lowest
poloidal and toroidal mode numbers, $m$ and $n$. High-$m$ tearing
modes were associated only with correspondingly high values of $\qRes
= m/n$ \cite{Furth73, White77, Hazeltine79, Goodall84}.
Recently it was demonstrated for the case of $\qRes=1$ double
and triple tearing modes that modes with high $m$ may dominate over
low-$m$ modes in a cylindrical plasma even on low-$q$ resonant
surfaces \cite{Bierwage05a}. There it was found that in configurations
where the distance between neighboring resonant surfaces is
sufficiently small, broad spectra of unstable modes exist and that the
dominant modes can have $m \sim \mathcal{O}(10)$.

Motivated by the results of Ref.~\cite{Bierwage05a}, in the present
work we investigate the conditions under which broad spectra of
multiple tearing modes with dominant higher-$m$ modes may arise. For
this purpose numerical simulations based on the reduced resistive
magnetohydrodynamic model in cylindrical geometry are used. The
dependence of the linear growth rate on parameters such as the
inter-resonance distance, the dissipation coefficients as well as the
magnetic shear will be addressed.

The main results of the present work are the following. It is shown
that broad spectra of unstable DTMs are also found in
configurations with $\qRes > 1$, thus generalizing the results
of Ref.~\cite{Bierwage05a}. The instability of modes with $m>1$
depends strongly on the distance between adjacent resonant
surfaces. For the resistivity dependence, a scaling law $\gLin \propto
\SHp^{-\alpha}$ is found with $1/3 \leq \alpha \leq 3/5$, which is
valid in a range of high values of $\SHp$. This is in agreement with
former studies \cite{Pritchett80, Ishii00}. For intermediate values of
$\SHp$ no power law is identified and for low $\SHp$ it is found that
$\gLin \propto \SHp$ independently of $m$. For high $m$ and small
$D_{12}$ the range where the relation $\gLin \propto \SHp^{-\alpha}$
does not hold can be rather wide, and it is especially in this regime
where high-$m$ modes are found to be dominant. This regime may include
ranges of $\SHp$ that are  relevant to tokamak operation.

The present paper is organized as follows. Section~\ref{sec:model}
contains an introduction of the model equations and the numerical
method used. Numerical results are presented in
Section~\ref{sec:results}, followed by a discussion and conclusions in
Section~\ref{sec:discussion}.

%===========================================
\section{Model}
\label{sec:model}

%-------------------------------------------
\subsection{Reduced MHD equations}

The reduced set of magnetohydrodynamic (RMHD) equations in cylindrical
geometry \cite{Strauss76} is used. It is obtained from the full MHD
model through a high-aspect ratio and low-$\beta$ expansion: $\epsilon
= a/R_0 \ll 1$, $\beta \ll 1$ (e.g., \cite{NishikawaWakatani}). Here,
$a$ and $R_0$ are, respectively, the minor and major radius of the
torus, $\epsilon$ is the inverse aspect ratio, and the plasma beta
$\beta = p/(B_0^2/2\mu_0)$, is the ratio of thermodynamic pressure
to magnetic pressure.

The high-aspect-ratio ordering and the presence of a strong axial
magnetic field $B_0$ allow to express the magnetic field in terms of
a magnetic flux function $\Psi$ as ${\bf B} = B_0\hat{\bf z} 
+ \nabla\Psi\times\hat{\bf z}$. Here, $\hat{\bf z}$ is the unit vector
in the axial direction and $B_0$ is taken to be constant. In the MHD
ordering, the single-fluid velocity ${\bf V}$ is approximated by the
${\bf E}\times{\bf B}$ drift velocity ${\bf V}_{E} =
-\nabla\Phi\times\hat{\bf z}/B_0$, where $\Phi$ is the electrostatic
potential (or stream function). In the limit of zero $\beta$ the
pressure equation is decoupled and the RMHD equations take the form of
a two-field model:
\begin{eqnarray}
\partial_t\Psi &=& -{\bf B}\cdot\nabla\Phi +
\frac{\eta}{\mu_0}\nabla_\perp^2\Psi + E_z,
\label{eq:rmhd01}
\\
\rho_{\rm m}{\rm d}_t \nabla_\perp^2\Phi &=& -\frac{1}{\mu_0}{\bf
B}\cdot\nabla\nabla_\perp^2\Psi + \nu_{\rm m}
\nabla_\perp^2\nabla_\perp^2\Phi.
\label{eq:rmhd02}
\end{eqnarray}

\noindent Here $\rho_{\rm m}$ is the mass density (constant due to the
assumption of incompressibility), $\eta$ the plasma resistivity and
$\nu_{\rm m}$ is the kinematic viscosity. The time-independent
electric field $E_z$ satisfies $\nabla E_z\times\hat{\bf z} = 0$. The
convective derivative is defined as ${\rm d}_t = \partial_t + {\bf
V}_{\rm E}\cdot\nabla$.

Cylindrical geometry is chosen, which gives the right-handed set of
coordinates $(r, \vartheta, z)$ where $r \in [0, a]$ is the radius,
$\vartheta \in [0,2\pi]$ the poloidal angle and $z \in [0,2\pi R_0]$
the axial  coordinate (related to the toroidal angle $\varphi$ via $z
= -R_0\varphi$). The Poisson bracket is defined as $[f,g] =
\frac{1}{r}(\partial_r f\partial_\vartheta g - \partial_r
g\partial_\vartheta f)$, and the Laplacian is approximated by
$\nabla^2 \approx \nabla_\perp^2 = \frac{1}{r}\partial_r r\partial_r +
\frac{1}{r^2}\partial_\vartheta^2$. In this ordering, toroidal
topology is preserved by retaining periodicity in the axial coordinate
$z$ and the poloidal angle $\vartheta$.

Normalizing the time by the poloidal Alfv\'{e}n time $\tHp =
\sqrt{\mu_0\rho_{\rm m}}a/B_{\vartheta}(a)$, the radial coordinate by
$a$, and introducing the angular coordinate $\zeta \equiv z/R_0 q(a)$,
the normalized variables are $\psi = \Psi/aB_\vartheta(a)$ and $\phi =
\Phi/(a^2/\tHp)$. Note that $\zeta$ is related to the toroidal angle
$\varphi$ via $\zeta = -\varphi/q(a)$, with $q(a) = aB_0/R_0
B_\vartheta(a)$ being the safety factor at $r=a$ [c.f.,
Eq.~(\ref{eq:qDef})]. The normalization for the source term is
$E_\zeta = E_z/[\eta_0 B_\vartheta(a)/\mu_0 a]$. With these
normalizations the Eqs.~(\ref{eq:rmhd01}) and (\ref{eq:rmhd02}) become
\begin{eqnarray}
\partial_t\psi &=& \left[\psi,\phi\right] - \partial_\zeta\phi -
\SHp^{-1}\left(\hat{\eta}j - E_\zeta\right),
\label{eq:rmhd1}
\\
\partial_t u &=& \left[u,\phi\right] + \left[j,\psi\right] +
\partial_\zeta j + \RHp^{-1}\nabla_\perp^2 u,
\label{eq:rmhd2}
\end{eqnarray}

\noindent where the current density $j$ and the vorticity $u$ are
related to $\psi$ and $\phi$ through $j = -\nabla_\perp^2\psi$ and $u
= \nabla_\perp^2\phi$, respectively. The magnetic Reynolds number
$\SHp$ (also called Lundquist number) and the kinematic Reynolds
number $\RHp$ are defined as $\SHp \equiv \tau_{\eta}/\tHp$ and $\RHp =
a^2/(\nu_{\rm m}\tHp/\rho_{\rm m})$, respectively. Here, $\tau_\eta =
a^2\mu_0/\eta(r=0)$ is the resistive diffusion time. The resistivity
profile is given by $\hat{\eta}(r)$, normalized such that
$\hat{\eta}(r=0) = 1$. The relative strengths of kinetic viscosity and
(resistive) diffusion are characterized by the Prandtl number $Pr =
\SHp/\RHp \propto \nu/\eta$, with $\nu = \nu_{\rm m}/\rho_{\rm m}$
being the specific ion viscosity.

%-------------------------------------------
\subsection{RMHD equilibrium and resonant modes}

The equilibrium state is defined as $\partial_t\psi = \partial_t\phi =
0$. Since the effect of the plasma pressure is neglected, the
structure of the equilibrium magnetic field ${\bf B}_{\rm eq}$ is
given in terms of the tokamak safety factor $q$, defined as
\begin{equation}
q = \frac{{\bf B}_{\rm eq}\cdot\nabla\zeta}{{\bf B}_{\rm
eq}\cdot\nabla\vartheta}.
\label{eq:qDef}
\end{equation}

\noindent The safety factor measures the field line pitch by counting
how many times a magnetic field line goes the long way around the
torus ($2\pi R_0$) after one turn the short way around ($2\pi
r$). Here, $q = q(r)$, so that the equilibrium magnetic flux surfaces,
where $\psi = {\rm const.}$, are uniquely defined by the radius
$r$. Each field variable $f$ is written in terms of a time-independent
equilibrium component $\overline{f}$ and a time-dependent perturbation
$\widetilde{f}$ as
\begin{equation}
f(r,\vartheta,\zeta,t) = \overline{f}(r) + \widetilde{f}(r, \vartheta,
\zeta, t).
\end{equation}

\noindent It is assumed that the equilibrium state is free of flows,
\begin{equation}
\eqPhi = \eqVor = 0.
\end{equation}

In general, a tokamak plasma has magnetic surfaces where $\qRes
= q(\rs) = m/n$, with integers $m$ and $n$. These are called rational
or resonant surfaces. The radius $\rs$ is called resonant radius,
because an infinitesimally small helical magnetic perturbation with
helicity $h = m/n$ is \emph{resonant} with the magnetic field
structure in the vicinity of $\rs$. Such a resonant perturbation
$\delta\flPsi_{m,n}$ does not bend field lines, so there is no
restoring force. If the perturbation leads the system to a state of
lower energy, the amplitude of the resonant perturbation will grow and
the mode is said to be unstable.

%-------------------------------------------
\subsection{Fourier representation}

In order to study the properties of such resonant modes it is useful
to apply a Fourier transform with respect to the periodic coordinates:
$(\vartheta, \zeta) \rightarrow (m,n)$. This representation also gives
an efficient and accurate numerical model. Substituting for each field
variable $f$ in Eqs.~(\ref{eq:rmhd1}) and (\ref{eq:rmhd2}) the Fourier
expansion
\begin{equation}
f(r, \vartheta, \zeta, t) = \frac{1}{2}\sum_{m,n}
f_{m,n}(r,t).e^{i(m\vartheta - n\zeta)} + {\rm c.c.},
\end{equation}

\noindent one obtains equations for the individual Fourier modes,
\begin{eqnarray}
\partial_t\psi_{m,n} &=& \left[\psi,\phi\right]_{m,n} + in\phi_{m,n}
\label{eq:rmhd1_mn}
\\
&& - \SHp^{-1}\left(\hat{\eta}j_{m,n}- E_{m,n}\right),
\nonumber
\\
\partial_t u_{m,n} &=& \left[u,\phi\right]_{m,n} +
\left[j,\psi\right]_{m,n} - in j_{m,n}
\label{eq:rmhd2_mn}
\\
&& + \RHp^{-1}\nabla_{m,n}^2 u_{m,n}, \nonumber
\end{eqnarray}

\noindent where $\nabla_{m,n}^2 = \frac{1}{r}\partial_r r\partial_r -
m^2/r^2$ and the nonlinear terms $[f,g]_{m,n}$ have acquired the form
of convolutions. In this model, if a perturbation is applied only to
modes of a given helicity $h=m/n$, modes with different helicities
will not be excited and the problem is effectively reduced to a
two-dimensional one. In the following, we will
exclusively refer to individual Fourier components of the field
variables and usually omit the $(m,n)$ subscripts for
convenience. Note that the equilibrium fields $\overline{f} =
\overline{f}(r)$ have only $(m,n) = (0,0)$ components.

%-------------------------------------------
\subsection{Equilibrium model}
\label{sec:equlib_model}

\begin{figure}
[tbp]
\centering
\caption{(Color online) Safety factor profiles with two resonant
surfaces used in this study. The Case (Ia), plotted in (a), has two
$\qRes = 1$ resonances, located at $r = \rsa$ and $\rsb$, a relatively
small distance $D_{12} = 0.06$ apart. In (b), variants of this
profile, with different inter-resonance distances $D_{12}$ and
different $\qRes$ are shown. The parameter values required to
reproduce Case (Ia) are given in Table~\protect\ref{tab:q-parm} and
the geometric characteristics of all cases are listed in
Table~\protect\ref{tab:equlib}.}
\label{fig:equlib_2tm}%
\end{figure}
% TEXTOR-94 (Ref.~\protect\cite{Koslowski97}, Fig.~10, \#68639

In order to study configurations with two resonant surfaces, the
following model formula for the equilibrium $q$ profile is used:
\begin{equation}
q(r) = q_0.F_1(r).\left\{1 +
\left(r/r_0\right)^{2\mu(r)}\right\}^{1/\mu(r)},
\label{eq:qModel}
\end{equation}

\noindent where
\begin{eqnarray}
r_0 &=& r_{\rm A}\left|\left[m/(nq_0)\right]^{\mu(r_{\rm A})} -
1\right|^{-1/2\mu(r_{\rm A})}, \nonumber
\\
\mu(r) &=& \mu_0 + \mu_1r^2, \nonumber
\\
F_1(r) &=& 1 + f_1\exp\left\{-\left[(r -
r_{11})/r_{12}\right]^2\right\}. \nonumber
\end{eqnarray}

\noindent The parameter set $\{q_0, r_{\rm A}, \mu_0, \mu_1\, m, n\}$
is used to design the underlying monotonic profile. The parameters
$\{f_1, r_{11}, r_{12}\}$ describe the Gaussian bump which is used
to create non-monotonic profiles with two (or three) $\qRes =
m/n$ resonant surfaces. The DTM configurations used in this study are
shown in Fig.~\ref{fig:equlib_2tm}. The corresponding model parameters
are listed in Table~\ref{tab:q-parm} and the geometric characteristics
of all cases are given in Table~\ref{tab:equlib}.

With $q$ given by Eq.~(\ref{eq:qModel}), $\eqPsi$ and $\eqCur$ were
calculated using the relations
\begin{equation}
q^{-1} = -\frac{1}{r}\frac{{\rm d}\eqPsi}{{\rm d}r} \quad {\rm and}
\quad \eqCur = -\nabla_\perp^2\eqPsi.
\label{eq:q-equlib}
\end{equation}

\noindent The resonant surfaces where $q=\qRes$ are labeled with
$\rsa$ and $\rsb$, and the inter-resonance distance is given by
$D_{12} = |\rsb - \rsa|$. The magnetic shear profile is defined as
\begin{equation}
s = \frac{r}{q} \frac{{\rm d}q}{{\rm d}r} = \frac{{\rm d}(\ln q)}{{\rm
d}(\ln r)},
\label{eq:s-equlib}
\end{equation}

\noindent and the local magnetic shear at the resonant radius $r_{{\rm
s}i}$ is $s_i = s(r_{{\rm s}i})$.

The time-independent source term $E_\zeta$ compensates the resistive
dissipation of the equilibrium current profile, i.e., $E_\zeta =
\hat{\eta}\eqCur$ (diffusive equilibrium). For numerical simulations
where the temporal evolution of the resistivity profile $\hat{\eta}$
is neglected one often assumes $\hat{\eta}(r) =
\eqCur(r=0)/\eqCur(r)$. This will generally lead to different values
of the resistivity at different $r_{{\rm s}i}$. In order to simplify
the comparison between growth rates of modes associated with different
resonant surfaces, a homogeneous resistivity profile, $\hat{\eta}(r) =
1$, is used in this study. Numerical tests indicate that the details
of $\hat{\eta}$ have no significant effect on any of the qualitative
characteristics discussed in this paper.

\begin{table}
[tbp]
\centering
\begin{ruledtabular}
\begin{tabular}[t]{ccccccccc}
$q_0$ & $r_{\rm A}$ & $\mu_0$ & $\mu_1$ & $m$ &
$n$ & $f_1$ & $r_{11}$ & $r_{12}$ \\
\hline $1.3$ & $0.655$ & $3.8824$ & $0$ & $1$ & $1$ & $-0.238$
& $0.4286$ & $0.304$ \\
\end{tabular}
\end{ruledtabular}
\caption{Parameter values for Eq.~(\protect\ref{eq:qModel}), giving
the $q$ profile shown in Fig.~\protect\ref{fig:equlib_2tm}(a). The
other cases are readily obtained by changing $m$ and $n$ (so that
$\qRes = m/n$) and $q_0$.}
\label{tab:q-parm}
\end{table}

\begin{table}
[tbp]
\centering
\begin{ruledtabular}
\begin{tabular}[t]{c|ccccc}
Case & $\qRes$ & $\qMin$ & $D_{12}$ & $s_1$ & $s_2$
\\
\hline (Ia) & $1$ & $0.99$ & $0.06$ & $-0.10$ & $0.12$ \\
(Ib) & $1$ & $0.96$ & $0.21$ & $-0.20$ & $0.45$ \\
(Ic) & $1$ & $0.92$ & $0.31$ & $-0.20$ & $0.66$ \\
(II) & $3/2$ & $1.49$ & $0.06$ & $-0.10$ & $0.12$ \\
(IIIa) & $2$ & $1.92$ & $0.31$ & $-0.20$ & $0.66$ \\
(IIIb) & $2$ & $1.99$ & $0.06$ & $-0.10$ & $0.12$ \\
(IV) & $5/2$ & $2.49$ & $0.06$ & $-0.10$ & $0.12$ \\
(V) & $3$ & $2.99$ & $0.06$ & $-0.10$ & $0.12$ \\
\end{tabular}
\end{ruledtabular}
\caption{Geometric properties of the $q$ profiles shown in
Fig.~\protect\ref{fig:equlib_2tm}(b).} 
\label{tab:equlib}
\end{table}

%-------------------------------------------
\subsection{Linearized equations}

When the amplitudes and gradients of the perturbed fields are
sufficiently small the nonlinear terms in Eqs.~(\ref{eq:rmhd1_mn}) and
(\ref{eq:rmhd2_mn}) may be neglected and one obtains the linearized
RMHD equations. In the linear system, the time dependence of a
perturbed field variable $\widetilde{f}$ takes the form
\begin{equation}
\widetilde{f}(r,t) = f(r).\exp(\lambda t),
\label{eq:t-lin}
\end{equation}

\noindent where $\lambda$ is a complex number.
Using Eq.~(\ref{eq:t-lin}), the system of equations
(\ref{eq:rmhd1_mn}) and (\ref{eq:rmhd2_mn}) becomes
\begin{eqnarray}
\lambda\flPsi &=& i\left(n-\frac{m}{q}\right) \flPhi +
\frac{1}{\SHp}\left(\laplR - \frac{m^2}{r^2}\right)\flPsi,
\label{eq:rmhd_lin1}
\\
\lambda\flVor &=& i\left(n-\frac{m}{q}\right)
\left(\frac{1}{r}\partial_r r\partial_r -
\frac{m^2}{r^2}\right)\flPsi
\label{eq:rmhd_lin2}
\\
&&\quad + \frac{im}{r}\left(\frac{s(s-2)}{rq} - \frac{\partial_r
s}{q}\right) \flPsi \nonumber
\\
&&\quad + \frac{1}{\RHp}\left(\laplR -
\frac{m^2}{r^2}\right)\flVor. \nonumber
\end{eqnarray}

\noindent Equations~(\ref{eq:rmhd_lin1}) and
(\ref{eq:rmhd_lin2}) are obtained by applying
Eqs.~(\ref{eq:q-equlib}) and (\ref{eq:s-equlib}) and expressing the
equilibrium fields in terms of the safety factor $q = q(r)$ and the
magnetic shear $s = s(r)$.

Modes for which the linear growth rate is positive, $\gLin =
\Re\{\lambda\} > 0$, are said to be linearly unstable. Their
amplitudes grow exponentially in time, e.g.,
\begin{equation}
\delta\flPsi_{m,n}(r,t) = \delta\flPsi_{m,n}(r)\exp({\gLin t})
\end{equation}

\noindent (in RMHD, $\Im\{\lambda\} = 0$). The linear growth rate
$\gLin$ is a function of the mode numbers,
\begin{equation}
\gLin = \gLin(m,n)
\end{equation}

\noindent (spectrum of growth rates), and written this way it is
generally referred to as the dispersion relation. Since this study is
restricted to modes with unique helicity, $h$, it is sufficient to
specify $m$ for a given $\qRes = h$ (so $n = m/\qRes$). Along with the
radial structure of the eigenmodes, $\flPsi(r)$
and $\flPhi(r)$, the linear growth rates are our main tool for
characterizing the linear instability of DTMs under various
conditions.

%-------------------------------------------
\subsection{Numerical method}

\begin{figure}
[tbp]
\centering
\caption{(Color online) Numerical convergence of the linear growth
rates $\gLin(m)$ with increasing $N_r$ (number of radial grid points
per unit length). The configuration used is Case (Ia) with $D_{12} =
0.06$, $\SHp = 10^8$ and $\RHp = 10^{12}$. The growth rates are shown
for modes $m=1,2,5$ and $8$. The $m=5$ mode is dominant in this
case. The small symbols represent EVP results, calculated in the range
$250 \leq N_r \leq 1430$. The large symbols represent IVP result. The
IVP solver was run with up to $N_r = 3000$. The agreement between both
methods and the numerical convergence is clearly shown. The horizontal
dotted lines indicate the linear growth rate values for $N_r =
730$. This is the standard $N_r$-value used for $\SHp = 10^8$ in this
paper (cf., Table~\ref{tab:S-Nr}). It can be seen that these growth
rates differ from the $N_r=3000$-results only by a few percent. This
accuracy is sufficient for the purpose of this paper.}
\label{fig:gr_convergence}%
\end{figure}

\begin{table}
[tbp]
\centering
\begin{ruledtabular}
\begin{tabular}[t]{c|c|c|c|c}
$\SHp$: & $\lesssim 10^5$ & $10^5 \rightarrow 10^7$ & $10^7
\rightarrow 10^8$ & $> 10^8$
\\
\hline $N_r$: & $180$--$360$ & $430$--$570$ & $730$ &
$1460$
\end{tabular}
\end{ruledtabular}
\caption{Typical values for $N_r$ (number of radial grid points per
unit length) used in the EVP solver. The IVP code is usually run with
$2000 \lesssim N_r \lesssim 3000$.}
\label{tab:S-Nr}
\end{table}

After discretizing Eqs.~(\ref{eq:rmhd_lin1}) and (\ref{eq:rmhd_lin2})
with respect to the radial coordinate, the linearized RMHD model may
be written as a generalized eigenvalue problem,
\begin{equation}
{\bf B}{\bf \Lambda}{\bf x} = {\bf A}{\bf x},
\end{equation}

\noindent where ${\bf x}^{\rm T} = \left[{\bf \flPsi},{\bf
\flPhi}\right]$, ${\bf \Lambda}$ is a diagonal matrix containing the
eigenvalues $\lambda$, and ${\bf A}$ and ${\bf B}$ are coefficient
matrices, which also include the finite-difference operators. All
eigenmodes $\textbf{x}_j$ with corresponding eigenvalues $\lambda_j$
are obtained using an eigenvalue problem (EVP) solver. The accuracy of
the results is checked by running tests with different numbers of
radial grid points and through comparison with results obtained by
solving the linearized version of Eqs.~(\ref{eq:rmhd01}) and
(\ref{eq:rmhd02}) as an initial value problem (IVP). The numerical
convergence and the agreement between the EVP and IVP results are
demonstrated in Fig.~\ref{fig:gr_convergence}. In Table~\ref{tab:S-Nr}
the radial resolution used for different values of $\SHp$ is
specified. Due to numerical constraints, the regime $\SHp > 10^{10}$
was not accessible with sufficient accuracy.

%===========================================
\section{Results}
\label{sec:results}

%-------------------------------------------
\subsection{Mode structures}
\label{sec:results_mstruc}

\begin{figure}
[tbp]
\centering
\caption{(Color online) Mode structures of the unstable $\qRes=1$ DTM
eigenmodes of Case (Ia), where $D_{12} = 0.06$. The locations of the
resonant surfaces are indicated by vertical dashed lines. Diagrams (a)
and (c) show, respectively, $\psi$ and $\phi/r$ of the $m=1$ mode. In
(b) and (d) the mode structures for $m=8$ are shown, which are
representative for other modes with $m>1$. In our notation, the
eigenmode $\Ma$ extends to $r=\rsa$, and $\Mb$ to $\rsb$. These
results are obtained with $\SHp = 10^6$, but similar mode structures
are also found for other values of $\SHp > 10^5$. The linear growth
rates $\gLin^{(1)}$ and $\gLin^{(2)}$ are also shown in (b) and (d).}
\label{fig:mstruc_2tm1_0.99_1e6_m18}%
\end{figure}

\begin{figure}
[tbp]
\centering
\caption{(Color online) Mode structures of the unstable $\qRes=2$ DTM
eigenmodes with $(m,n) = (2,1)$. In (a) and (c), respectively, $\psi$
and $\phi/r$ are shown for Case (IIIa) ($D_{12} = 0.31$). In (b) and
(d) the results for Case (IIIb) ($D_{12} = 0.06$) are plotted. These
results were obtained with $\SHp = 10^6$.}
\label{fig:mstruc_2tm2_0.92-0.99_1e6_m2}%
\end{figure}

The radial structure of an eigenmode determines
which part of the plasma is affected by the instability, since
$\phi(r)$ is related to the radial and poloidal components of the
plasma displacement velocity,
\begin{equation}
v_r = -\frac{im}{r}\phi \quad {\rm and} \quad v_\vartheta =
\partial_r\phi.
\end{equation}

In general, for each $(m,n)$ there may be as many unstable eigenmodes
as there are resonant surfaces $\qRes = m/n$. The radial mode
structures of unstable $\qRes=1$ DTMs are shown in
Fig.~\ref{fig:mstruc_2tm1_0.99_1e6_m18}. The configuration used is
Case (Ia), where $D_{12} = 0.06$. For $m=1$
[Fig.~\ref{fig:mstruc_2tm1_0.99_1e6_m18}(a) and (c)] there are two
unstable eigenmodes. For $m>1$
[Fig.~\ref{fig:mstruc_2tm1_0.99_1e6_m18}(b) and (d)] only one
eigenmode is unstable (the $m=8$ mode structure shown is
representative for other modes with $m > 1$). In our notation, the
eigenmode $\Ma$ is associated with the innermost resonant surface
$\rsa$. For $m=1$ it has a finite amplitude in the region $0 < r <
\rsa$. The eigenmode $\Mb$ is active in the region $0 < r < \rsb$ for
$m=1$, and mainly in the region $\rsa < 0 < \rsb$ for $m > 1$.

In Figure~\ref{fig:mstruc_2tm2_0.92-0.99_1e6_m2}, the mode structures
of $\qRes=2$ DTMs with $(m,n) = (2,1)$ are plotted. Results are
shown for two cases with different distances $D_{12}$: Case (IIIa)
with $D_{12} = 0.31$ [Fig.~\ref{fig:mstruc_2tm2_0.92-0.99_1e6_m2}(a)
and (c)], and Case (IIIb) with $D_{12} = 0.06$
[Fig.~\ref{fig:mstruc_2tm2_0.92-0.99_1e6_m2}(b) and (d)]. In
agreement with previous works on DTMs, it is found that two individual
modes are present in the case of larger $D_{12}$: $\Ma$ with even
parity around $\rsa$ and $\Mb$ with odd parity \cite{Kerner82,
Ishii00}. In the case of smaller $D_{12}$ only the eigenmode $\Mb$ is
found to be unstable.

The $\Ma$ eigenmode is essentially an STM associated with the $r=\rsa$
resonant surface (here, a negative-shear surface), since it is
practically unaffected by the presence of resonant surfaces beyond $r
= \rsa$. The actual DTMs (for sufficiently small $D_{12}$) are the
$\Mb$ eigenmodes. The radial structure of DTMs with $m>1$ is very
similar for different $\qRes$, as may be seen by comparing the
profiles (b) and (d) in Figs.~\ref{fig:mstruc_2tm1_0.99_1e6_m18} and
\ref{fig:mstruc_2tm2_0.92-0.99_1e6_m2}. Let us note that Cases (II),
(IV) and (V) of Table~\ref{tab:equlib}, where $\qRes = 3/2$,
$5/2$ and $3$, respectively, have eigenmode structures very similar to
those found for $\qRes = 2$ in
Fig.~\ref{fig:mstruc_2tm2_0.92-0.99_1e6_m2}.

The local resistivity $\eta(r_{{\rm s}i})$ and the magnetic shear
$s(r_{{\rm s}i})$ at a resonant surface $r_{{\rm s}i}$, together with
the distance between neighboring resonant surfaces $D_{12}$, determine
which eigenmode, $\Ma$ or $\Mb$, will be the dominant mode for a given 
$m$. An exception is the STM-like eigenmode $\Ma$: as its mode
structure indicates, it is not affected by $D_{12}$.

%-------------------------------------------
\subsection{Viscosity effect}
\label{sec:results_nu}

\begin{figure}
[tbp]
\centering
\caption{(Color online) Dependence of the linear DTM growth rates on
the Prandtl number [Case (Ia)]. The growth rates of eigenmodes with
$m=1,2,8$ are plotted for fixed $\SHp = 10^6$.}
\label{fig:scale-nu_2tm1_0.99_m128}%
\end{figure}

Although, it is difficult to obtain reliable values for the ion
viscosity in the tokamak core, it is usually thought to be very
low. However, it is required in most nonlinear simulations for the
purpose of providing a cut-off at short wavelengths in order to be
consistent with the finite number of grid points or Fourier modes. As
may be seen from Fig.~\ref{fig:scale-nu_2tm1_0.99_m128}, as long as
the Prandtl number satisfies $Pr \lesssim 0.1$, the viscosity has
practically no effect on the linear growth rates of DTMs. A similar
result is also obtained for other values of $\SHp$ as well as for
single and triple tearing modes, so it may be considered a generic
characteristic of tearing modes. The mode structures in
Figs.~\ref{fig:mstruc_2tm1_0.99_1e6_m18} and
\ref{fig:mstruc_2tm2_0.92-0.99_1e6_m2} and all the following results
were obtained in the regime where $Pr < 0.01$, so the viscosity effect
will not be discussed further in this paper. Let us note that the
stabilizing effect of viscosity for $Pr \gtrsim 1$, evident in
Fig.~\ref{fig:scale-nu_2tm1_0.99_m128}, was pointed out previously by
Ofman \cite{Ofman92} in a study on DTMs in the presence of equilibrium
shear flows.

%-------------------------------------------
\subsection{Dispersion relations for $\qRes \geq 1$}
\label{sec:results_qres}

\begin{figure}
[tbp]
\centering
\caption{(Color online) Comparison between dispersion relations for
(a) $\qRes = 1$, $2$ and $3$, and (b) $\qRes = 3/2$ and $5/2$. For all
cases $D_{12} = 0.06$ (cf., Table~\protect\ref{tab:equlib}). For
$m=1$, circles in diagram (a), growth rates of both unstable modes are
shown: $\gLin^{(1)} = 4.1\times 10^{-3}$, $\gLin^{(2)} = 1.6\times
10^{-3}$. All other growth rates belong to $\Mb$-type eigenmodes,
since the $\Ma$-type modes are stable for $m>1$ in these cases. For
all cases $\SHp = 10^6$.}
\label{fig:spec_qres}%
\end{figure}

Recently it was discovered that configurations with nearby $\qRes = 1$
resonant surfaces have broad spectra of linearly unstable modes
\cite{Bierwage05a}. In particular, if the distance between the
resonant radii is sufficiently small, modes with $m \sim
\mathcal{O}(10)$ were found to have linear growth rates several times
higher than the $m=1$ mode.

Indeed, similar behavior is observed when the resonances are located
at higher $q$ values, such as $\qRes=3/2$, $2$, $5/2$ and $3$,
corresponding to Cases (IIb), (III), (IV) and (V), respectively. The
dispersion relations for these cases are
shown in Fig.~\ref{fig:spec_qres}. All these $q$ profiles have the
same values for $D_{12}$, $s_1$ and $s_2$. They only differ from each
other in the values of $\qMin$ and $\qRes$. Let us remark
that for all cases the width of the spectrum as well as the location
of the maximum growth rate is practically identical: $\gLin > 0$ for
$1 \leq m \lesssim 18$, and the peak
\begin{equation}
\gPeak = \gLin(\mPeak) = {\rm Max}\{\gLin(m)\}
\label{eq:spec_peak}
\end{equation}

\noindent is located at $m=8$. Except for the $m=1$ mode [$\qRes
= 1$, Case (Ia)], all growth rates in Fig.~\ref{fig:spec_qres} are
associated with $\Mb$-type eigenmodes. Due to these results it may be
conjectured that all DTMs [except for $\Mb(m=1)$] behave similarly for
any $\qRes$, including $\qRes = 1$.

%-------------------------------------------
\subsection{Role of $D_{12}$}
\label{sec:results_dij}

\begin{figure}
[tbp]
\centering
\caption{(Color online) Role of $D_{12}$ (variable shears) for the
instability of $\qRes = 1$ DTMs. In (a) the $q$ profiles are shown
[Cases (Ia,b,c) in Table~\protect\ref{tab:equlib}] and in (b) the
corresponding dispersion relations are plotted. For all cases $\SHp =
10^6$, and only the growth rate of the dominant mode is shown for each
$m$.}
\label{fig:q-spec_2tm_qmin}%
\end{figure}

\begin{figure}
[tbp]
\centering
\caption{(Color online) Role of $D_{12}$ (constant shears) for the
instability of $\qRes = 1$ DTMs. In (a) the $q$ profiles used are
shown. The basic profile is Case (Ib), with $D_{12} = 0.21$. The other
profiles ($D_{12} \leq 0.17$) were obtained by gradually reducing
$D_{12}$, but holding the shears $s_1 = -0.20$ and $s_2 = 0.45$
constant. For this, the model equation (\protect\ref{eq:qModel}) was
multiplied by another factor, $F_2(r)$, defined equivalently to
$F_1(r)$. In (b) the corresponding dispersion relations are shown. For
all cases $\SHp = 10^6$, and only the growth rate of the dominant mode
is shown for each $m$.}
\label{fig:q-spec_2tm_dr}%
\end{figure}

\begin{figure}
[tbp]
\centering
\caption{(Color online) $D_{12}$ dependence of the growth rates of
$m=1$ and $m=2$ DTMs with $\qRes = 1$. While varying $D_{12}$, the
shears $s_1 = -0.20$ and $s_2 = 0.45$ were held constant. The values
for Lundquist number are (a) $\SHp = 10^6$ and (b) $\SHp = 10^7$. For
several data points, indicated by arrows, estimates are given for the
linear resistive layer width $\delta_\eta$ defined by
Eq.~(\protect\ref{eq:delta}).}
\label{fig:g1-dr_const-s_1e6-1e7}%
\end{figure}

Consider the scenario where $\qMin$ gradually drops below $\qRes=1$
due to an increase in the current density. During
this process the distance between the resonant surfaces will grow in
time, starting from $D_{12} = 0$ when $\qMin = \qRes$. The series of
Cases (Ia), (Ib) and (Ic), shown in
Fig.~\ref{fig:q-spec_2tm_qmin} (a), represent the equilibria at
successive instants in time during such a process. Similar cases were
previously investigated in Ref.~\cite{Ishii00} where the effect of
$D_{12}$ on $\gLin(m=3,n=1)$ and the corresponding scaling exponent
$\alpha$ (in $\gLin \propto \SHp^{-\alpha}$) was characterized. In
Ref.~\cite{Sato01}, the role of $\qMin$ was studied in the
context of partial and full reconnection.

The dispersion relations for the three cases of
Fig.~\ref{fig:q-spec_2tm_qmin}(a) are plotted in
Fig.~\ref{fig:q-spec_2tm_qmin}(b). Clearly, the width of the spectrum
is reduced as $D_{12}$ increases, and the $m=1$ mode eventually
becomes dominant.

Note that Cases (Ia)--(Ic) all have different magnetic shears $s_1$
and $s_2$ [cf., Table~\ref{tab:equlib}]. The effect of varying only
$D_{12}$ is illustrated in Fig.~\ref{fig:q-spec_2tm_dr}. Starting with
Case (Ib), where $D_{12} = 0.21$, the $q$ profile is gradually
modified in such a way that $D_{12}$ decreases down to 0.08, while
both shears, $s_1$ and $s_2$, are held constant. The profiles used are
plotted in Fig.~\ref{fig:q-spec_2tm_dr}(a) and the corresponding
dispersion relations in Fig.~\ref{fig:q-spec_2tm_dr}(b). These
results show that $D_{12}$ controls the broadness of the spectrum.
Modes with higher $m$ and with higher growth rates than the lower-$m$
modes appear when $D_{12}$ is decreased.

It is known that a tearing mode is stable when the local magnetic
shear at the resonant surface is zero \cite{Furth63}. The results in
Fig.~\ref{fig:q-spec_2tm_qmin} indicate that for DTMs with $m>1$ the
destabilizing effect of small $D_{12}$ dominates over the effect of
low local shears [$|s_1|$ and $s_2$ become small when $\Delta q =
\qRes - \qMin$ is reduced (cf., Table~\ref{tab:equlib})].

The dependence of the DTM growth rates on the parameter $\qMin$
(meaning, simultaneous variation of $D_{12}$, $s_1$ and $s_2$) was
studied previously by Ishii \textit{et al.} \cite{Ishii00} for the
$(m,n)=(3,1)$ mode and a set of $q$ profiles similar to those in
Fig.~\ref{fig:q-spec_2tm_qmin}(a). They found that the curve
$\gLin(\qMin)$ is not monotonic. In
Fig.~\ref{fig:g1-dr_const-s_1e6-1e7}(a) it is shown
that this is also true for the dependence of $\gLin$ on
$D_{12}$ alone (constant shears): in the range shown ($0.08 \leq
D_{12} \leq 0.21$), the growth rate of the $\Mb(m=1)$ mode,
$\gLin^{(2)}(m=1)$, has a maximum around $D_{12} = 0.14$. A comparison
between the results for $\SHp = 10^6$ in
Fig.~\ref{fig:g1-dr_const-s_1e6-1e7}(a) and $\SHp = 10^7$ in
Fig.~\ref{fig:g1-dr_const-s_1e6-1e7}(b) shows that the value $\SHp$
influences the location of the peak. This is most likely related to
the distance $D_{12}$ becoming comparable to the linear resistive
layer width $\delta_\eta$, which is estimated after
Ref.~\cite{Pritchett80},
\begin{equation}
\delta_\eta \simeq \left[\frac{\gLin(m)}{(m/r_{\rm min})^2 B_{\rm
s}^{'2}\SHp}\right]^{1/4},
\label{eq:delta}
\end{equation}

\noindent where $B_{\rm s}' = s(r_{{\rm s}i})/q(r_{{\rm s}i})$. In the
following section, the role of $\SHp$ will be studied in more detail.

It is noted that the growth rate $\gLin^{(1)}(m=1)$ does not depend on
$D_{12}$, as it is expected from the mode structure $\Ma(m=1)$
[Fig.~\ref{fig:mstruc_2tm1_0.99_1e6_m18}(a) and (c)]. Hence, the
increase of the $m=1$ growth rate in Fig.~\ref{fig:q-spec_2tm_qmin}(b)
is caused by the increase in the magnetic shear $|s_1|$.

%-------------------------------------------
\subsection{Role of resistivity}
\label{sec:results_eta}

\begin{figure}
[tbp]
\centering
\caption{(Color online) DTM dispersion relations for $\SHp = 10^6$,
$10^7$ and $10^8$ obtained with Case (Ia) where $\qRes = 1$. For $m=1$
there are two unstable eigenmodes, $\Ma$ and $\Mb$. The growth rates
of both the dominant ($\Ma$) and the secondary ($\Mb$) $m=1$ eigenmode
are shown for $\SHp = 10^6$ (they almost coincide for higher $\SHp$).
Vertical dashed lines indicate the locations $\mPeak$ of the peaks
$\gPeak = {\rm Max}\{\gLin(m)\}$. Similar results are
obtained for cases with $\qRes > 1$.}
\label{fig:spec_eta}%
\end{figure}

In the previous sections the linear instability of DTMs was
investigated at $\SHp = 10^6$, since this value lies in the regime
where  nonlinear simulations of MHD instabilities are often performed:
$10^4 \lesssim \SHp \lesssim 10^7$. While large tokamaks typically
operate in regimes where $\SHp \gtrsim 10^8$ (except in the early
current-ramp-up phase), it is difficult to access this low-collisional
regime with nonlinear simulations. Therefore, the linear study of the
$\SHp$ dependence of DTM growth rates is an important tool for
relating nonlinear simulation results to plasma conditions.

In Fig.~\ref{fig:spec_eta} the dispersion relations of Case (Ia) is
shown for $\SHp = 10^6$, $10^7$ and $10^8$. The growth rates of all
modes are found to decrease with increasing $\SHp$, as it is expected
for resistive instabilities. Let us remark that the width of the
spectrum is not affected by the variation of $\SHp$:
unstable modes are found in the range $1 \leq m \leq 18$. However, it
can be seen that the growth rates of modes with higher $m$ drop more
rapidly than growth rates of low-$m$ modes, when $\SHp$ is
increased, which is also reflected by the shift of the peak $\gPeak$
to lower $m$. This is important, because the mode number of the
fastest-growing mode determines the size of the magnetic islands
formed in the early nonlinear regime \cite{Bierwage05a}.

In the remaining part of this section, the $\SHp$ dependence of DTM
growth rates is examined in detail. Comparisons between low- and
higher-$m$ modes are made and the role of the distance $D_{12}$ will
also be emphasized. First, cases where $\qRes = 1$ are
considered, whereby a distinction is made
between DTMs with $m=1$ and $m>1$. The results obtained for $\qRes =
1$ are readily applied to cases with $\qRes > 1$, as will
be shown at the end of this section where $\qRes = 2$ DTMs are
considered.

%...........................................
\subsubsection{DTMs with $m=1$}

\begin{figure}
[tbp]
\centering
\caption{(Color online) $\SHp$ dependence of the linear growth rates
of the two $m=1$ DTM eigenmodes $\Ma$ and $\Mb$ of Case (Ia) [cf.,
Fig.~\protect\ref{fig:mstruc_2tm1_0.99_1e6_m18}, (a) and (c)]. For
comparison the growth-rate scalings of ``corresponding'' STMs are
shown as well. This means, that the instability of the DTM eigenmode
$\Ma$ is compared with the STM mode on the resonant surface $(\rsa,
s_1)$ (monotonic profile with negative shear), while $\Mb$ is compared
with the STM mode on $(\rsb, s_2)$ (positive-shear profile). The STM
results are plotted as straight lines, fitted to the actual data. In
the regime $\SHp \gtrsim 10^6$, the $\gamma_{\rm lin} \propto
\SHp^{-\alpha}$ scaling with $\alpha = 1/3$ is followed closely by
both STMs: $\alpha \approx 0.32$.}
\label{fig:scale-eta_2tm1-1tm_0.99}%
\end{figure}

\begin{figure}
[tbp]
\centering
\caption{(Color online) $\SHp$ dependence of the linear growth rates
of the two $m=1$ DTM eigenmodes, (a) for $D_{12} = 0.31$ [Case (Ic)]
and (b) for $D_{12} = 0.21$ [Case (Ib)]. Corresponding STM data are
shown as well, as in Fig.~\protect\ref{fig:scale-eta_2tm1-1tm_0.99}.}
\label{fig:scale-eta_2tm1-1tm_0.92-0.96}%
\end{figure}

It is known that DTMs with $m=1$ behave similarly to $m=1$ STMs,
regardless of the location of the resonant surfaces and their mutual
distance \cite{Furth73}. Due to this close relationship between the
$m=1$ STMs and the $m=1$ DTMs it is possible to decompose the
non-monotonic $q$ profile into two monotonic ones --- one with a
resonance at $\rsa$ and negative shear $s_1$, the other with a
resonance at $\rsb$ and positive shear $s_2$ --- and compare the
growth rates and $\SHp$ dependences of the STM eigenmodes with the
corresponding DTMs. Our aim is to characterize deviations between STMs
and DTMs in certain ranges of $\SHp$. For clarity, both $\Ma$- and
$\Mb$-type eigenmodes in a double-tearing configuration are referred
to as ``DTMs.''

In Fig.~\ref{fig:scale-eta_2tm1-1tm_0.99} this comparison is
performed for Case (Ia), where $D_{12} = 0.06$
[Fig.~\ref{fig:equlib_2tm} (a)]. The results for Cases (Ib) and (Ic),
where $D_{12} = 0.21$ and $D_{13} = 0.31$, respectively, are presented
in Fig.~\ref{fig:scale-eta_2tm1-1tm_0.92-0.96}.

In Case (Ia), $|s_1| \approx s_2$ (cf., Table~\ref{tab:equlib}), so
that the growth rates of the two STMs are almost equal
(Fig.~\ref{fig:scale-eta_2tm1-1tm_0.99}). Both have $\alpha = 0.32
\approx 1/3$, in agreement with linear theory. It can be observed that
the DTM growth rates coincide with the corresponding STM growth rates
only at relatively high values of $\SHp$: $\gLin^{(1)}$ for $\SHp >
10^7$ and $\gLin^{(2)}$ for $\SHp > 10^8$. The agreement between $m=1$
DTMs and STMs for intermediate $\SHp$ is significantly improved when 
the distance between the resonant surfaces is increased. This can be
seen by comparing Fig.~\ref{fig:scale-eta_2tm1-1tm_0.99} with
Fig.~\ref{fig:scale-eta_2tm1-1tm_0.92-0.96}.

The linear growth rate of a tearing mode plotted as a function of the
magnetic Reynolds number, $\gLin(\SHp)$, always has a maximum at a
certain value $\SHp = \Smax$. For instance, for $\gLin^{(2)}$ in
Fig.~\ref{fig:scale-eta_2tm1-1tm_0.99}, one finds $\Smax \approx
5\times 10^5$. In the regime where $\SHp < \Smax$ the width of the
linear resistive layer $\delta_\eta$ [Eq.~(\ref{eq:delta})] is
comparable to $D_{12}$. Due to the strong resistive diffusion in the
regime $\SHp \lesssim \Smax$, the two resonant surfaces are
effectively seen as a single $\qRes=1$ ``surface.'' This leads
to the reduction in the DTM growth rate apparent in
Fig.~\ref{fig:scale-eta_2tm1-1tm_0.99}. For $\SHp \ll \Smax$ a linear
dependence $\gLin \propto \SHp$ (i.e., $\alpha = -1$) is obtained
independently of $m$.

In summary, for $m=1$ DTMs it is found that for sufficiently small
$D_{12}$, the growth rate of the $\Mb$ mode exhibits no scaling law in
a wide range of $\SHp$. For $\SHp \ll \Smax$ a linear dependence is
observed and for $\SHp \gg \Smax$ the scaling exponent $\alpha = 1/3$
is obtained for both eigenmodes. However, the range of $\SHp$ in which
deviations from the $\alpha = 1/3$ scaling are observed may extend to
high $\SHp$ when $D_{12}$ is small.

%...........................................
\subsubsection{DTMs with $m>1$}

\begin{figure}
[tbp]
\centering
\caption{(Color online) $\SHp$ dependence of the linear growth rates
of $\qRes = 1$ DTMs with $m=1,2,3$ for a small distance $D_{12} =
0.06$ [Case (Ia)]. The dotted lines represent the scaling law $\gLin =
\SHp^{-\alpha}$ with $\alpha = 1/3$ and $3/5$.}
\label{fig:scale-eta_2tm1_0.99_m128}%
\end{figure}

\begin{figure}
[tbp]
\centering
\caption{(Color online) $\SHp$ dependence of the linear growth rates
of $\qRes = 1$ DTMs with $m=1,2$ for the distance $D_{12} = 0.21$
[Case (Ib)].}
\label{fig:scale-eta_2tm1_0.96_m12}%
\end{figure}

Here the $\SHp$ dependence of $\qRes = 1$ DTMs with $m>1$ is
analyzed. First, consider Case (Ia) where the inter-resonance distance
is small: $D_{12} = 0.06$. As indicated in Fig.~\ref{fig:spec_eta}, we
have selected the growth rates $\gLin^{(2)}(m=2)$ and
$\gLin^{(2)}(m=8)$, and plotted them as functions of $\SHp$ in
Fig.~\ref{fig:scale-eta_2tm1_0.99_m128}. It is noted that in Case (Ia)
only $\Mb$-type modes are unstable for $m>1$. For comparison,
$\gLin^{(1)}(m=1)$ and $\gLin^{(2)}(m=1)$ from
Fig.~\ref{fig:scale-eta_2tm1-1tm_0.99} are shown as well. It can be
seen that, similarly to the $m=1$ mode, $\gLin^{(2)}(m=2)$ and
$\gLin^{(2)}(m=8)$ have $\alpha = -1$ for $\SHp \ll \Smax$, where
$\Smax(m=2) \approx 10^5$ and $\Smax(m=8) \approx 2\times 10^5$. In
the range of $\SHp$ shown in Fig.~\ref{fig:scale-eta_2tm1_0.99_m128},
only $\gLin^{(2)}(m=8)$ approaches the scaling $\alpha = 3/5$ in the
limit of high $\SHp$. The growth rate of the $m=2$ mode,
$\gLin^{(2)}(m=2)$, has $\alpha$ close to but somewhat larger than
$1/3$. Calculations with $m=4$ and $m=6$ (not shown here) gave
intermediate values $1/3 < \alpha_m < 3/5$ (with $\alpha_m$ being the
scaling exponent for the mode number $m$ in the limit of $\SHp \gg
\Smax$).

In Fig.~\ref{fig:scale-eta_2tm1_0.96_m12} the growth rates of the
$m=1$ and $m=2$ eigenmodes are compared for Case (Ib), where $D_{12} =
0.21$. In contrast to Case (Ia)
[Fig.~\ref{fig:scale-eta_2tm1_0.99_m128}, $D_{12} = 0.06$], now there
are two unstable $m=2$ eigenmodes. A scaling exponent $\alpha_{m=2}
\approx 3/5$ is obtained for both $m=2$ eigenmodes [while
$\alpha_{m=2} \approx 1/3$ in Case (Ia)].

As a result of the complicated dependence of $\gLin$ on the parameter
set $\{m, s_1, s_2, D_{12}, \SHp\}$, growth rates $\gLin(m>1)$ may
rise above $\gLin(m=1)$ in certain regimes of the parameter space.
This can be seen in Fig.~\ref{fig:scale-eta_2tm1_0.99_m128}:
$\gLin^{(2)}(m=8) > \gLin^{(1)}(m=1)$ for $4\times 10^4 \lesssim \SHp
\lesssim 2\times 10^8$, and $\gLin^{(2)}(m=2) > \gLin^{(1)}(m=1)$ for
$\SHp \gtrsim 8\times 10^5$ (upper limit not known). In this regime
dispersion relations are found to peak at $\mPeak > 1$.

%...........................................
\subsubsection{Cases with $\qRes = 2$}

\begin{figure}
[tbp]
\centering
\caption{(Color online) $\SHp$ dependence of the linear growth rates
of $\qRes = 2$ DTMs. In diagram (a) the growth rates of the two $m=2$
eigenmodes of Case (IIIa) with $D_{12} = 0.31$ are shown [cf.,
Fig.~\protect\ref{fig:mstruc_2tm2_0.92-0.99_1e6_m2}(a) and (c)]. In
(b) the growth rates of the unstable $m=2$ and $m=8$ eigenmodes (both
$\Mb$-type) of Case (IIIb) are plotted [cf.,
Fig.~\protect\ref{fig:mstruc_2tm2_0.92-0.99_1e6_m2}(b) and (d)].}
\label{fig:scale-eta_2tm2_0.92-0.99_m28}%
\end{figure}

In Fig.~\ref{fig:scale-eta_2tm2_0.92-0.99_m28} the $\SHp$ dependence
of the linear growth rates of $\qRes = 2$ DTMs is shown. A case
with relatively large $D_{12}$ is plotted in
Fig.~\ref{fig:scale-eta_2tm2_0.92-0.99_m28}(a). Both $m=2$ eigenmodes
follow the scaling law $\gLin \propto \SHp^{-3/5}$ rather well. When
$D_{12}$ is reduced the growth-rate scalings deviate from this power
law in a wide range of $\SHp$, as can be seen in
Fig.~\ref{fig:scale-eta_2tm2_0.92-0.99_m28}(b). While the higher-$m$
mode still approaches $\alpha_m = 3/5$, the lower-$m$ mode has
$\alpha_m$ close to $1/3$. Note the similarity between
Fig.~\ref{fig:scale-eta_2tm2_0.92-0.99_m28} and the $\qRes = 1$
results shown above.

%===========================================
\section{Discussion and Conclusions}
\label{sec:discussion}

Double tearing modes with high poloidal mode numbers are destabilized
when the distance between the resonant surfaces is small. For a given
inter-resonance distance the mode number of the fastest growing mode
was observed to shift to lower $m$ when $\SHp$ is increased
(Fig.~\ref{fig:spec_eta}). This is related to the fact that modes with
different mode numbers $m$ approach scaling laws $\gLin \propto
\SHp^{-\alpha}$ with different exponents $\alpha = \alpha_{\rm
m}$. Moreover, and in agreement with earlier works \cite{Pritchett80,
Ishii00}, it was found that the scaling exponent $\alpha_m$ is a
function of the inter-resonance distance $D_{12}$ (e.g.,
Figs.~\ref{fig:scale-eta_2tm1_0.99_m128} and
\ref{fig:scale-eta_2tm1_0.96_m12}).

Linear tearing mode theory predicts two characteristic values for the
$\SHp$-scaling exponent: $\alpha = 1/3$ for $m=1$ modes \cite{Coppi76}
and DTMs on nearby resonant surfaces \cite{Pritchett80}, and $\alpha =
3/5$ for STMs \cite{Furth63}. When the distance $D_{12}$ is increased,
DTMs are transformed into STMs and $\alpha_m$ gradually increases from
$1/3$ to $3/5$ (except for $m=1$ modes) \cite{Ishii00}. In the present
work, it was observed that the transition from $\alpha_m = 1/3$ to $3/5$
also occurs when $m$ is increased
(Figs.~\ref{fig:scale-eta_2tm1_0.99_m128},
\ref{fig:scale-eta_2tm1_0.96_m12} and
\ref{fig:scale-eta_2tm2_0.92-0.99_m28}). However, the variation of $m$
does not change the character of the DTM mode structures, which is in
contrast with the above-mentioned effect of increasing $D_{12}$
(Figs.~\ref{fig:mstruc_2tm1_0.99_1e6_m18} and
\ref{fig:mstruc_2tm2_0.92-0.99_1e6_m2}).

For practical reasons, the knowledge of the mode number of the
dominant mode $\mPeak$ in the spectrum $\gLin(m)$ is important, since
it determines the structure of the magnetic islands in the early
nonlinear regime \cite{Bierwage05a}.
The observations that $\mPeak$ varies with $\SHp$ and that the
growth rates $\gLin(m)$ may approach their respective characteristic
scaling exponent $\alpha_m$ only at very high values of $\SHp$
[c.f., Figs.~\ref{fig:scale-eta_2tm1_0.99_m128},
\ref{fig:scale-eta_2tm1_0.96_m12} and
\ref{fig:scale-eta_2tm2_0.92-0.99_m28}(b)] have the following
important consequence.
Results obtained from nonlinear simulations run in the collisional
regime (e.g., $\SHp \sim 10^6$) for a given configuration may not be
easily extrapolated to higher-$\SHp$, since magnetic islands with
different poloidal mode numbers are expected for different
$\SHp$. Note, however, that the effective magnetic Reynolds number in
the reconnection regions may be smaller due to the action of
micro-turbulence (anomalous resistivity) \cite{Ji98,
BreslauJardin03}.

In summary, the linear instability characteristics of DTMs with high
poloidal mode numbers $m$ were studied numerically. High-$m$ tearing
modes become unstable when two or more resonant surfaces $q(\rsa) =
q(\rsb) = \qRes$ are formed in a tokamak plasma and when the
distance between these resonances $D_{12} = |\rsb - \rsa|$ is still
small. It was shown that despite the low magnetic shear in the
vicinity of $\qMin$, modes with high $m$ have high growth rates due to
the destabilizing effect of small $D_{12}$. The width of the DTM
spectrum and mode number of the dominant mode were found to be
independent of the $q$-value. Broad spectra of unstable DTMs, with
dominant modes having $m>1$, were found in a wide range of magnetic
Reynolds numbers, including the regimes in which tokamaks operate.
Let us note that this result may also be applied to configurations
with more than two resonant surfaces \cite{Bierwage05a} and
configurations with low magnetic shear \cite{Kleva87}.

The findings of this linear study motivate a nonlinear investigation
of DTMs. Nonlinear simulations were performed in the past for
relatively large inter-resonance distances (e.g., \cite{Carreras79,
Chang96, Ishii00, Sato01}). According to our results, during the stage
where the inter-resonance distance is still small, fast growing
high-$m$ DTMs may significantly modify the $q$ profile near $\qMin$
and thereby affect the long-term evolution \cite{Bierwage05a}. This
might have important implications for the understanding of the
sawtooth crash \cite{Porcelli96, Hastie98, Itoh98} and other
applications of DTM dynamics mentioned in the introduction, such as
the anomalous current penetration during the current-ramp up phase,
%\cite{Schmidt71, Furth73, Stix76},
off-axis sawteeth
%\cite{Chang96}
or the formation of ITBs.
%\cite{ITB04}.

An effect that was neglected here but may decouple DTMs in practice
is differential rotation \cite{Ofman92, Persson94, Shen98}. However,
its influence decreases with decreasing inter-resonance distance. A
factor that is expected to be important in regions with low magnetic
shear near resonant surfaces (as in the vicinity of $\qMin$) is the
pressure gradient \cite{Waelbroeck88}. This and other extensions are
left for future study.

%-------------------------------------------
\begin{acknowledgements}
A.B. would like to thank Y. Kishimoto for valuable
discussions. S.B. acknowledges the Graduate School of Energy Science
at Kyoto University for its support and hospitality.

This work is partially supported by the 21st Century COE Program at
Kyoto University.
\end{acknowledgements}

%===========================================


\begin{thebibliography}{45}
\expandafter\ifx\csname natexlab\endcsname\relax\def\natexlab#1{#1}\fi
\expandafter\ifx\csname bibnamefont\endcsname\relax
  \def\bibnamefont#1{#1}\fi
\expandafter\ifx\csname bibfnamefont\endcsname\relax
  \def\bibfnamefont#1{#1}\fi
\expandafter\ifx\csname citenamefont\endcsname\relax
  \def\citenamefont#1{#1}\fi
\expandafter\ifx\csname url\endcsname\relax
  \def\url#1{\texttt{#1}}\fi
\expandafter\ifx\csname urlprefix\endcsname\relax\def\urlprefix{URL }\fi
\providecommand{\bibinfo}[2]{#2}
\providecommand{\eprint}[2][]{\url{#2}}

\bibitem[{\citenamefont{White et~al.}(1977)\citenamefont{White, Monticello,
  Rosenbluth, and Waddell}}]{White77}
\bibinfo{author}{\bibfnamefont{R.~B.} \bibnamefont{White}},
  \bibinfo{author}{\bibfnamefont{D.~A.} \bibnamefont{Monticello}},
  \bibinfo{author}{\bibfnamefont{M.~N.} \bibnamefont{Rosenbluth}},
  \bibnamefont{and} \bibinfo{author}{\bibfnamefont{B.~V.}
  \bibnamefont{Waddell}}, in \emph{\bibinfo{booktitle}{Proceedings of the
  Conference on Plasma Physics and Controlled Nuclear Fusion Research,
  Berchtesgaden, Germany, 1976}} (\bibinfo{publisher}{International Atomic
  Energy Agency}, \bibinfo{address}{Vienna}, \bibinfo{year}{1977}),
  vol.~\bibinfo{volume}{1}, p. \bibinfo{pages}{569}.

\bibitem[{\citenamefont{Pritchett et~al.}(1980)\citenamefont{Pritchett, Lee,
  and Drake}}]{Pritchett80}
\bibinfo{author}{\bibfnamefont{P.~L.} \bibnamefont{Pritchett}},
  \bibinfo{author}{\bibfnamefont{Y.~C.} \bibnamefont{Lee}}, \bibnamefont{and}
  \bibinfo{author}{\bibfnamefont{J.~F.} \bibnamefont{Drake}},
  \bibinfo{journal}{Phys. Fluids} \textbf{\bibinfo{volume}{23}},
  \bibinfo{pages}{1368} (\bibinfo{year}{1980}).

\bibitem[{\citenamefont{Mahajan and Hazeltine}(1982)}]{Mahajan82}
\bibinfo{author}{\bibfnamefont{S.~M.} \bibnamefont{Mahajan}} \bibnamefont{and}
  \bibinfo{author}{\bibfnamefont{R.~D.} \bibnamefont{Hazeltine}},
  \bibinfo{journal}{Nucl. Fusion} \textbf{\bibinfo{volume}{22}},
  \bibinfo{pages}{1191} (\bibinfo{year}{1982}).

\bibitem[{\citenamefont{Bierwage et~al.}(2005)\citenamefont{Bierwage,
  Hamaguchi, Wakatani, Benkadda, and Leoncini}}]{Bierwage05a}
\bibinfo{author}{\bibfnamefont{A.}~\bibnamefont{Bierwage}},
  \bibinfo{author}{\bibfnamefont{S.}~\bibnamefont{Hamaguchi}},
  \bibinfo{author}{\bibfnamefont{M.}~\bibnamefont{Wakatani}},
  \bibinfo{author}{\bibfnamefont{S.}~\bibnamefont{Benkadda}}, \bibnamefont{and}
  \bibinfo{author}{\bibfnamefont{X.}~\bibnamefont{Leoncini}},
  \bibinfo{journal}{Phys. Rev. Lett.} \textbf{\bibinfo{volume}{94}},
  \bibinfo{pages}{065001} (\bibinfo{year}{2005}).

\bibitem[{\citenamefont{Schmidt and Yoshikawa}(1971)}]{Schmidt71}
\bibinfo{author}{\bibfnamefont{J.}~\bibnamefont{Schmidt}} \bibnamefont{and}
  \bibinfo{author}{\bibfnamefont{S.}~\bibnamefont{Yoshikawa}},
  \bibinfo{journal}{Phys. Rev. Lett.} \textbf{\bibinfo{volume}{26}},
  \bibinfo{pages}{753} (\bibinfo{year}{1971}).

\bibitem[{\citenamefont{Furth et~al.}(1973)\citenamefont{Furth, Rutherford, and
  Selberg}}]{Furth73}
\bibinfo{author}{\bibfnamefont{H.~P.} \bibnamefont{Furth}},
  \bibinfo{author}{\bibfnamefont{P.~H.} \bibnamefont{Rutherford}},
  \bibnamefont{and} \bibinfo{author}{\bibfnamefont{H.}~\bibnamefont{Selberg}},
  \bibinfo{journal}{Phys. Fluids} \textbf{\bibinfo{volume}{16}},
  \bibinfo{pages}{1054} (\bibinfo{year}{1973}).

\bibitem[{\citenamefont{Stix}(1976)}]{Stix76}
\bibinfo{author}{\bibfnamefont{T.~H.} \bibnamefont{Stix}},
  \bibinfo{journal}{Phys. Rev. Lett.} \textbf{\bibinfo{volume}{36}},
  \bibinfo{pages}{521} (\bibinfo{year}{1976}).

\bibitem[{\citenamefont{Goodall and Wesson}(1984)}]{Goodall84}
\bibinfo{author}{\bibfnamefont{D.~H.~J.} \bibnamefont{Goodall}}
  \bibnamefont{and} \bibinfo{author}{\bibfnamefont{J.~A.}
  \bibnamefont{Wesson}}, \bibinfo{journal}{Plasma Phys. Control. Nucl. Fusion}
  \textbf{\bibinfo{volume}{26}}, \bibinfo{pages}{789} (\bibinfo{year}{1984}).

\bibitem[{\citenamefont{Edwards et~al.}(1986)\citenamefont{Edwards, Campbell,
  Engelhardt, Fahrbach, Gill, Granetz, Tsuji, Tubbing, Weller, Wesson
  et~al.}}]{Edwards86}
\bibinfo{author}{\bibfnamefont{A.~W.} \bibnamefont{Edwards}},
  \bibinfo{author}{\bibfnamefont{D.~J.} \bibnamefont{Campbell}},
  \bibinfo{author}{\bibfnamefont{W.~W.} \bibnamefont{Engelhardt}},
  \bibinfo{author}{\bibfnamefont{H.-U.} \bibnamefont{Fahrbach}},
  \bibinfo{author}{\bibfnamefont{R.~D.} \bibnamefont{Gill}},
  \bibinfo{author}{\bibfnamefont{R.~S.} \bibnamefont{Granetz}},
  \bibinfo{author}{\bibfnamefont{S.}~\bibnamefont{Tsuji}},
  \bibinfo{author}{\bibfnamefont{B.~J.~D.} \bibnamefont{Tubbing}},
  \bibinfo{author}{\bibfnamefont{A.}~\bibnamefont{Weller}},
  \bibinfo{author}{\bibfnamefont{J.}~\bibnamefont{Wesson}},
  \bibnamefont{et~al.}, \bibinfo{journal}{Phys. Rev. Lett.}
  \textbf{\bibinfo{volume}{57}}, \bibinfo{pages}{210} (\bibinfo{year}{1986}).

\bibitem[{\citenamefont{Taylor et~al.}(1986)\citenamefont{Taylor, Efthimion,
  Arunasalam, Goldston, Grek, Hill, Johnson, McGuire, Ramsey, and
  Stauffer}}]{Taylor86}
\bibinfo{author}{\bibfnamefont{G.}~\bibnamefont{Taylor}},
  \bibinfo{author}{\bibfnamefont{P.~C.} \bibnamefont{Efthimion}},
  \bibinfo{author}{\bibfnamefont{V.}~\bibnamefont{Arunasalam}},
  \bibinfo{author}{\bibfnamefont{R.~J.} \bibnamefont{Goldston}},
  \bibinfo{author}{\bibfnamefont{B.}~\bibnamefont{Grek}},
  \bibinfo{author}{\bibfnamefont{K.~W.} \bibnamefont{Hill}},
  \bibinfo{author}{\bibfnamefont{D.~W.} \bibnamefont{Johnson}},
  \bibinfo{author}{\bibfnamefont{K.}~\bibnamefont{McGuire}},
  \bibinfo{author}{\bibfnamefont{A.~T.} \bibnamefont{Ramsey}},
  \bibnamefont{and} \bibinfo{author}{\bibfnamefont{F.~J.}
  \bibnamefont{Stauffer}}, \bibinfo{journal}{Nucl. Fusion}
  \textbf{\bibinfo{volume}{26}}, \bibinfo{pages}{339} (\bibinfo{year}{1986}).

\bibitem[{\citenamefont{Campbell et~al.}(1986)\citenamefont{Campbell, Gill,
  Gowers, Wesson, Bartlett, Best, Coda, Costley, Edwards, Kissel
  et~al.}}]{Campbell86}
\bibinfo{author}{\bibfnamefont{D.~J.} \bibnamefont{Campbell}},
  \bibinfo{author}{\bibfnamefont{R.~D.} \bibnamefont{Gill}},
  \bibinfo{author}{\bibfnamefont{C.~W.} \bibnamefont{Gowers}},
  \bibinfo{author}{\bibfnamefont{J.~A.} \bibnamefont{Wesson}},
  \bibinfo{author}{\bibfnamefont{D.~V.} \bibnamefont{Bartlett}},
  \bibinfo{author}{\bibfnamefont{C.~H.} \bibnamefont{Best}},
  \bibinfo{author}{\bibfnamefont{S.}~\bibnamefont{Coda}},
  \bibinfo{author}{\bibfnamefont{A.~E.} \bibnamefont{Costley}},
  \bibinfo{author}{\bibfnamefont{A.}~\bibnamefont{Edwards}},
  \bibinfo{author}{\bibfnamefont{S.~E.} \bibnamefont{Kissel}},
  \bibnamefont{et~al.}, \bibinfo{journal}{Nucl. Fusion}
  \textbf{\bibinfo{volume}{26}}, \bibinfo{pages}{1085} (\bibinfo{year}{1986}).

\bibitem[{\citenamefont{Kim}(1986)}]{Kim86}
\bibinfo{author}{\bibfnamefont{S.~B.} \bibnamefont{Kim}},
  \bibinfo{journal}{Nucl. Fusion} \textbf{\bibinfo{volume}{26}},
  \bibinfo{pages}{1251} (\bibinfo{year}{1986}).

\bibitem[{\citenamefont{Ishida et~al.}(1988)\citenamefont{Ishida, Shirai,
  Nakashima, Nishitani, Fukuda, and Team}}]{Ishida88}
\bibinfo{author}{\bibfnamefont{S.}~\bibnamefont{Ishida}},
  \bibinfo{author}{\bibfnamefont{H.}~\bibnamefont{Shirai}},
  \bibinfo{author}{\bibfnamefont{K.}~\bibnamefont{Nakashima}},
  \bibinfo{author}{\bibfnamefont{T.}~\bibnamefont{Nishitani}},
  \bibinfo{author}{\bibfnamefont{T.}~\bibnamefont{Fukuda}}, \bibnamefont{and}
  \bibinfo{author}{\bibfnamefont{J.-.} \bibnamefont{Team}},
  \bibinfo{journal}{Plasma Phys. Control. Fusion}
  \textbf{\bibinfo{volume}{30}}, \bibinfo{pages}{1069} (\bibinfo{year}{1988}).

\bibitem[{\citenamefont{Carreras et~al.}(1979)\citenamefont{Carreras, Hicks,
  and Waddell}}]{Carreras79}
\bibinfo{author}{\bibfnamefont{B.}~\bibnamefont{Carreras}},
  \bibinfo{author}{\bibfnamefont{H.~R.} \bibnamefont{Hicks}}, \bibnamefont{and}
  \bibinfo{author}{\bibfnamefont{B.~V.} \bibnamefont{Waddell}},
  \bibinfo{journal}{Nucl. Fusion} \textbf{\bibinfo{volume}{19}},
  \bibinfo{pages}{583} (\bibinfo{year}{1979}).

\bibitem[{\citenamefont{Parail and Pereverzev}(1980)}]{Parail80}
\bibinfo{author}{\bibfnamefont{V.~V.} \bibnamefont{Parail}} \bibnamefont{and}
  \bibinfo{author}{\bibfnamefont{G.~V.} \bibnamefont{Pereverzev}},
  \bibinfo{journal}{Sov. J. Plasma Phys.} \textbf{\bibinfo{volume}{6}},
  \bibinfo{pages}{14} (\bibinfo{year}{1980}).

\bibitem[{\citenamefont{Pfeiffer}(1985)}]{Pfeiffer85}
\bibinfo{author}{\bibfnamefont{W.}~\bibnamefont{Pfeiffer}},
  \bibinfo{journal}{Nucl. Fusion} \textbf{\bibinfo{volume}{25}},
  \bibinfo{pages}{673} (\bibinfo{year}{1985}).

\bibitem[{\citenamefont{Chang et~al.}(1996)\citenamefont{Chang, Park,
  Frederickson, Batha, Bell, Bell, Budny, Bush, Janos, Levinton
  et~al.}}]{Chang96}
\bibinfo{author}{\bibfnamefont{Z.}~\bibnamefont{Chang}},
  \bibinfo{author}{\bibfnamefont{W.}~\bibnamefont{Park}},
  \bibinfo{author}{\bibfnamefont{E.~D.} \bibnamefont{Frederickson}},
  \bibinfo{author}{\bibfnamefont{S.~H.} \bibnamefont{Batha}},
  \bibinfo{author}{\bibfnamefont{M.~G.} \bibnamefont{Bell}},
  \bibinfo{author}{\bibfnamefont{R.}~\bibnamefont{Bell}},
  \bibinfo{author}{\bibfnamefont{R.~V.} \bibnamefont{Budny}},
  \bibinfo{author}{\bibfnamefont{C.~E.} \bibnamefont{Bush}},
  \bibinfo{author}{\bibfnamefont{A.}~\bibnamefont{Janos}},
  \bibinfo{author}{\bibfnamefont{F.~M.} \bibnamefont{Levinton}},
  \bibnamefont{et~al.}, \bibinfo{journal}{Phys. Rev. Lett.}
  \textbf{\bibinfo{volume}{77}}, \bibinfo{pages}{3553} (\bibinfo{year}{1996}).

\bibitem[{\citenamefont{Persson and Dewar}(1994)}]{Persson94}
\bibinfo{author}{\bibfnamefont{M.}~\bibnamefont{Persson}} \bibnamefont{and}
  \bibinfo{author}{\bibfnamefont{R.~L.} \bibnamefont{Dewar}},
  \bibinfo{journal}{Phys. Plasmas} \textbf{\bibinfo{volume}{1}},
  \bibinfo{pages}{1256} (\bibinfo{year}{1994}).

\bibitem[{\citenamefont{Ishii et~al.}(2002)\citenamefont{Ishii, Azumi, and
  Kishimoto}}]{Ishii02}
\bibinfo{author}{\bibfnamefont{Y.}~\bibnamefont{Ishii}},
  \bibinfo{author}{\bibfnamefont{M.}~\bibnamefont{Azumi}}, \bibnamefont{and}
  \bibinfo{author}{\bibfnamefont{Y.}~\bibnamefont{Kishimoto}},
  \bibinfo{journal}{Phys. Rev. Lett.} \textbf{\bibinfo{volume}{89}},
  \bibinfo{pages}{205002} (\bibinfo{year}{2002}).

\bibitem[{\citenamefont{Waddell et~al.}(1979)\citenamefont{Waddell, Carreras,
  Hicks, and Holmes}}]{Waddell79}
\bibinfo{author}{\bibfnamefont{B.~V.} \bibnamefont{Waddell}},
  \bibinfo{author}{\bibfnamefont{B.}~\bibnamefont{Carreras}},
  \bibinfo{author}{\bibfnamefont{H.~R.} \bibnamefont{Hicks}}, \bibnamefont{and}
  \bibinfo{author}{\bibfnamefont{J.~A.} \bibnamefont{Holmes}},
  \bibinfo{journal}{Phys. Fluids} \textbf{\bibinfo{volume}{22}},
  \bibinfo{pages}{896} (\bibinfo{year}{1979}).

\bibitem[{\citenamefont{Connor et~al.}(1988)\citenamefont{Connor, Cowley,
  Hastie, Hender, Hood, and Martin}}]{Connor88}
\bibinfo{author}{\bibfnamefont{J.~W.} \bibnamefont{Connor}},
  \bibinfo{author}{\bibfnamefont{S.~C.} \bibnamefont{Cowley}},
  \bibinfo{author}{\bibfnamefont{R.~J.} \bibnamefont{Hastie}},
  \bibinfo{author}{\bibfnamefont{T.~C.} \bibnamefont{Hender}},
  \bibinfo{author}{\bibfnamefont{A.}~\bibnamefont{Hood}}, \bibnamefont{and}
  \bibinfo{author}{\bibfnamefont{T.~J.} \bibnamefont{Martin}},
  \bibinfo{journal}{Phys. Fluids} \textbf{\bibinfo{volume}{31}},
  \bibinfo{pages}{577} (\bibinfo{year}{1988}).

\bibitem[{\citenamefont{G\"{u}nter et~al.}(2000)\citenamefont{G\"{u}nter,
  Schade, Maraschek, Pinches, Strumberger, Wolf, Yu, and the {ASDEX}~{U}pgrade
  {T}eam}}]{Guenter00}
\bibinfo{author}{\bibfnamefont{S.}~\bibnamefont{G\"{u}nter}},
  \bibinfo{author}{\bibfnamefont{S.}~\bibnamefont{Schade}},
  \bibinfo{author}{\bibfnamefont{M.}~\bibnamefont{Maraschek}},
  \bibinfo{author}{\bibfnamefont{S.~D.} \bibnamefont{Pinches}},
  \bibinfo{author}{\bibfnamefont{E.}~\bibnamefont{Strumberger}},
  \bibinfo{author}{\bibfnamefont{R.}~\bibnamefont{Wolf}},
  \bibinfo{author}{\bibfnamefont{Q.}~\bibnamefont{Yu}}, \bibnamefont{and}
  \bibinfo{author}{\bibnamefont{the {ASDEX}~{U}pgrade {T}eam}},
  \bibinfo{journal}{Nucl. Fusion} \textbf{\bibinfo{volume}{40}},
  \bibinfo{pages}{1541} (\bibinfo{year}{2000}).

\bibitem[{\citenamefont{Voitsekhovitch
  et~al.}(2002)\citenamefont{Voitsekhovitch, Garbet, Benkadda, Beyer, and
  Figarella}}]{Voitsekhovitch02}
\bibinfo{author}{\bibfnamefont{I.}~\bibnamefont{Voitsekhovitch}},
  \bibinfo{author}{\bibfnamefont{X.}~\bibnamefont{Garbet}},
  \bibinfo{author}{\bibfnamefont{S.}~\bibnamefont{Benkadda}},
  \bibinfo{author}{\bibfnamefont{P.}~\bibnamefont{Beyer}}, \bibnamefont{and}
  \bibinfo{author}{\bibfnamefont{C.~F.} \bibnamefont{Figarella}},
  \bibinfo{journal}{Phys. Plasmas} \textbf{\bibinfo{volume}{9}},
  \bibinfo{pages}{4671} (\bibinfo{year}{2002}).

\bibitem[{\citenamefont{Connor et~al.}(2004)\citenamefont{Connor, Fukuda,
  Garbet, Gormezano, Mukhavotov, Wakatani, the {ITB}~{D}atabase Group, and the
  {T}opical {G}roup on {T}ransport {and} {I}nternal {T}ransport~{B}arrier
  {P}hysics}}]{Connor04}
\bibinfo{author}{\bibfnamefont{J.~W.} \bibnamefont{Connor}},
  \bibinfo{author}{\bibfnamefont{T.}~\bibnamefont{Fukuda}},
  \bibinfo{author}{\bibfnamefont{X.}~\bibnamefont{Garbet}},
  \bibinfo{author}{\bibfnamefont{C.}~\bibnamefont{Gormezano}},
  \bibinfo{author}{\bibfnamefont{V.}~\bibnamefont{Mukhavotov}},
  \bibinfo{author}{\bibfnamefont{M.}~\bibnamefont{Wakatani}},
  \bibinfo{author}{\bibnamefont{the {ITB}~{D}atabase Group}}, \bibnamefont{and}
  \bibinfo{author}{\bibnamefont{the {T}opical {G}roup on {T}ransport {and}
  {I}nternal {T}ransport~{B}arrier {P}hysics}}, \bibinfo{journal}{Nuclear
  Fusion} \textbf{\bibinfo{volume}{44}}, \bibinfo{pages}{R1}
  (\bibinfo{year}{2004}).

\bibitem[{\citenamefont{Coppi et~al.}(1976)\citenamefont{Coppi, Galvao, Pellat,
  Rosenbluth, and Rutherford}}]{Coppi76}
\bibinfo{author}{\bibfnamefont{B.}~\bibnamefont{Coppi}},
  \bibinfo{author}{\bibfnamefont{R.}~\bibnamefont{Galvao}},
  \bibinfo{author}{\bibfnamefont{R.}~\bibnamefont{Pellat}},
  \bibinfo{author}{\bibfnamefont{M.~N.} \bibnamefont{Rosenbluth}},
  \bibnamefont{and} \bibinfo{author}{\bibfnamefont{P.~H.}
  \bibnamefont{Rutherford}}, \bibinfo{journal}{Fiz. Plazmy}
  \textbf{\bibinfo{volume}{2}}, \bibinfo{pages}{961} (\bibinfo{year}{1976}),
  \bibinfo{note}{[Sov. J. Plasma Phys. \textbf{2}, 533 (1976)]}.

\bibitem[{\citenamefont{Furth et~al.}(1963)\citenamefont{Furth, Killeen, and
  Rosenbluth}}]{Furth63}
\bibinfo{author}{\bibfnamefont{H.~P.} \bibnamefont{Furth}},
  \bibinfo{author}{\bibfnamefont{J.}~\bibnamefont{Killeen}}, \bibnamefont{and}
  \bibinfo{author}{\bibfnamefont{M.~N.} \bibnamefont{Rosenbluth}},
  \bibinfo{journal}{Phys. Fluids} \textbf{\bibinfo{volume}{6}},
  \bibinfo{pages}{459} (\bibinfo{year}{1963}).

\bibitem[{\citenamefont{Yu}(1996)}]{Yu96}
\bibinfo{author}{\bibfnamefont{Q.}~\bibnamefont{Yu}}, \bibinfo{journal}{Phys.
  Plasmas} \textbf{\bibinfo{volume}{3}}, \bibinfo{pages}{2898}
  (\bibinfo{year}{1996}).

\bibitem[{\citenamefont{Ishii et~al.}(2000)\citenamefont{Ishii, Azumi, Kurita,
  and Tuda}}]{Ishii00}
\bibinfo{author}{\bibfnamefont{Y.}~\bibnamefont{Ishii}},
  \bibinfo{author}{\bibfnamefont{M.}~\bibnamefont{Azumi}},
  \bibinfo{author}{\bibfnamefont{G.}~\bibnamefont{Kurita}}, \bibnamefont{and}
  \bibinfo{author}{\bibfnamefont{T.}~\bibnamefont{Tuda}},
  \bibinfo{journal}{Phys. Plasmas} \textbf{\bibinfo{volume}{7}},
  \bibinfo{pages}{4477} (\bibinfo{year}{2000}).

\bibitem[{\citenamefont{Dong et~al.}(2003)\citenamefont{Dong, Mahajan, and
  Horton}}]{Dong03}
\bibinfo{author}{\bibfnamefont{J.~Q.} \bibnamefont{Dong}},
  \bibinfo{author}{\bibfnamefont{S.~M.} \bibnamefont{Mahajan}},
  \bibnamefont{and} \bibinfo{author}{\bibfnamefont{W.}~\bibnamefont{Horton}},
  \bibinfo{journal}{Phys. Plasmas} \textbf{\bibinfo{volume}{10}},
  \bibinfo{pages}{3151} (\bibinfo{year}{2003}).

\bibitem[{\citenamefont{Held et~al.}(1999)\citenamefont{Held, Leboeuf, and
  Carreras}}]{Held99}
\bibinfo{author}{\bibfnamefont{E.~D.} \bibnamefont{Held}},
  \bibinfo{author}{\bibfnamefont{J.~N.} \bibnamefont{Leboeuf}},
  \bibnamefont{and} \bibinfo{author}{\bibfnamefont{B.~A.}
  \bibnamefont{Carreras}}, \bibinfo{journal}{Phys. Plasmas}
  \textbf{\bibinfo{volume}{6}}, \bibinfo{pages}{837} (\bibinfo{year}{1999}).

\bibitem[{\citenamefont{Ofman}(1992)}]{Ofman92}
\bibinfo{author}{\bibfnamefont{L.}~\bibnamefont{Ofman}},
  \bibinfo{journal}{Phys. Fluids B} \textbf{\bibinfo{volume}{4}},
  \bibinfo{pages}{2751} (\bibinfo{year}{1992}).

\bibitem[{\citenamefont{Shen and Liu}(1998)}]{Shen98}
\bibinfo{author}{\bibfnamefont{C.}~\bibnamefont{Shen}} \bibnamefont{and}
  \bibinfo{author}{\bibfnamefont{Z.~X.} \bibnamefont{Liu}},
  \bibinfo{journal}{Plasma Phys. Control. Fusion}
  \textbf{\bibinfo{volume}{40}}, \bibinfo{pages}{1} (\bibinfo{year}{1998}).

\bibitem[{\citenamefont{G\"{u}nter et~al.}(1999)\citenamefont{G\"{u}nter,
  Giruzzi, Gude, Haye, Lackner, Maraschek, Schade, Sesnic, Wolf, Yu
  et~al.}}]{Guenter99}
\bibinfo{author}{\bibfnamefont{S.}~\bibnamefont{G\"{u}nter}},
  \bibinfo{author}{\bibfnamefont{G.}~\bibnamefont{Giruzzi}},
  \bibinfo{author}{\bibfnamefont{A.}~\bibnamefont{Gude}},
  \bibinfo{author}{\bibfnamefont{R.~J.~L.} \bibnamefont{Haye}},
  \bibinfo{author}{\bibfnamefont{K.}~\bibnamefont{Lackner}},
  \bibinfo{author}{\bibfnamefont{M.}~\bibnamefont{Maraschek}},
  \bibinfo{author}{\bibfnamefont{S.}~\bibnamefont{Schade}},
  \bibinfo{author}{\bibfnamefont{S.}~\bibnamefont{Sesnic}},
  \bibinfo{author}{\bibfnamefont{R.}~\bibnamefont{Wolf}},
  \bibinfo{author}{\bibfnamefont{Q.}~\bibnamefont{Yu}}, \bibnamefont{et~al.},
  \bibinfo{journal}{Plasma Phys. Control. Fusion}
  \textbf{\bibinfo{volume}{41}}, \bibinfo{pages}{B231} (\bibinfo{year}{1999}).

\bibitem[{\citenamefont{Hazeltine et~al.}(1979)\citenamefont{Hazeltine,
  Strauss, Mahajan, and Ross}}]{Hazeltine79}
\bibinfo{author}{\bibfnamefont{R.~D.} \bibnamefont{Hazeltine}},
  \bibinfo{author}{\bibfnamefont{H.~R.} \bibnamefont{Strauss}},
  \bibinfo{author}{\bibfnamefont{S.~M.} \bibnamefont{Mahajan}},
  \bibnamefont{and} \bibinfo{author}{\bibfnamefont{D.~W.} \bibnamefont{Ross}},
  \bibinfo{journal}{Phys. Fluids} \textbf{\bibinfo{volume}{22}},
  \bibinfo{pages}{1932} (\bibinfo{year}{1979}).

\bibitem[{\citenamefont{Strauss}(1976)}]{Strauss76}
\bibinfo{author}{\bibfnamefont{H.~R.} \bibnamefont{Strauss}},
  \bibinfo{journal}{Phys. Fluids} \textbf{\bibinfo{volume}{19}},
  \bibinfo{pages}{134} (\bibinfo{year}{1976}).

\bibitem[{\citenamefont{Nishikawa and Wakatani}(2000)}]{NishikawaWakatani}
\bibinfo{author}{\bibfnamefont{K.}~\bibnamefont{Nishikawa}} \bibnamefont{and}
  \bibinfo{author}{\bibfnamefont{M.}~\bibnamefont{Wakatani}},
  \emph{\bibinfo{title}{Plasma Physics}} (\bibinfo{publisher}{Springer,
  Berlin}, \bibinfo{year}{2000}).

\bibitem[{\citenamefont{Kerner and Tasso}(1982)}]{Kerner82}
\bibinfo{author}{\bibfnamefont{W.}~\bibnamefont{Kerner}} \bibnamefont{and}
  \bibinfo{author}{\bibfnamefont{H.}~\bibnamefont{Tasso}},
  \bibinfo{journal}{Plasma Phys.} \textbf{\bibinfo{volume}{24}},
  \bibinfo{pages}{97} (\bibinfo{year}{1982}).

\bibitem[{\citenamefont{Sato et~al.}(2001)\citenamefont{Sato, Hamaguchi, and
  Wakatani}}]{Sato01}
\bibinfo{author}{\bibfnamefont{M.}~\bibnamefont{Sato}},
  \bibinfo{author}{\bibfnamefont{S.}~\bibnamefont{Hamaguchi}},
  \bibnamefont{and} \bibinfo{author}{\bibfnamefont{M.}~\bibnamefont{Wakatani}},
  \bibinfo{journal}{J. Phys. Soc. Jpn.} \textbf{\bibinfo{volume}{70}},
  \bibinfo{pages}{2578} (\bibinfo{year}{2001}).

\bibitem[{\citenamefont{Ji et~al.}(1998)\citenamefont{Ji, Yamada, Hsu, and
  Kulsrud}}]{Ji98}
\bibinfo{author}{\bibfnamefont{H.}~\bibnamefont{Ji}},
  \bibinfo{author}{\bibfnamefont{M.}~\bibnamefont{Yamada}},
  \bibinfo{author}{\bibfnamefont{S.}~\bibnamefont{Hsu}}, \bibnamefont{and}
  \bibinfo{author}{\bibfnamefont{R.}~\bibnamefont{Kulsrud}},
  \bibinfo{journal}{Phys. Rev. Lett.} \textbf{\bibinfo{volume}{80}},
  \bibinfo{pages}{3256} (\bibinfo{year}{1998}).

\bibitem[{\citenamefont{Breslau and Jardin}(2003)}]{BreslauJardin03}
\bibinfo{author}{\bibfnamefont{J.~A.} \bibnamefont{Breslau}} \bibnamefont{and}
  \bibinfo{author}{\bibfnamefont{S.~C.} \bibnamefont{Jardin}},
  \bibinfo{journal}{Phys. Plasmas} \textbf{\bibinfo{volume}{10}},
  \bibinfo{pages}{1291} (\bibinfo{year}{2003}).

\bibitem[{\citenamefont{Kleva et~al.}(1987)\citenamefont{Kleva, Drake, and
  Denton}}]{Kleva87}
\bibinfo{author}{\bibfnamefont{R.~G.} \bibnamefont{Kleva}},
  \bibinfo{author}{\bibfnamefont{J.~F.} \bibnamefont{Drake}}, \bibnamefont{and}
  \bibinfo{author}{\bibfnamefont{R.~E.} \bibnamefont{Denton}},
  \bibinfo{journal}{Phys. Fluids} \textbf{\bibinfo{volume}{30}},
  \bibinfo{pages}{2119} (\bibinfo{year}{1987}).

\bibitem[{\citenamefont{Porcelli et~al.}(1996)\citenamefont{Porcelli, Boucher,
  and Rosenbluth}}]{Porcelli96}
\bibinfo{author}{\bibfnamefont{F.}~\bibnamefont{Porcelli}},
  \bibinfo{author}{\bibfnamefont{D.}~\bibnamefont{Boucher}}, \bibnamefont{and}
  \bibinfo{author}{\bibfnamefont{M.~N.} \bibnamefont{Rosenbluth}},
  \bibinfo{journal}{Plasma Phys. Control. Fusion}
  \textbf{\bibinfo{volume}{38}}, \bibinfo{pages}{2163} (\bibinfo{year}{1996}).

\bibitem[{\citenamefont{Hastie}(1998)}]{Hastie98}
\bibinfo{author}{\bibfnamefont{R.~J.} \bibnamefont{Hastie}},
  \bibinfo{journal}{Astrophys. Space Sci.} \textbf{\bibinfo{volume}{256}},
  \bibinfo{pages}{177} (\bibinfo{year}{1998}).

\bibitem[{\citenamefont{Itoh et~al.}(1998)\citenamefont{Itoh, Itoh, Zushi, and
  Fukuyama}}]{Itoh98}
\bibinfo{author}{\bibfnamefont{S.-I.} \bibnamefont{Itoh}},
  \bibinfo{author}{\bibfnamefont{K.}~\bibnamefont{Itoh}},
  \bibinfo{author}{\bibfnamefont{H.}~\bibnamefont{Zushi}}, \bibnamefont{and}
  \bibinfo{author}{\bibfnamefont{A.}~\bibnamefont{Fukuyama}},
  \bibinfo{journal}{Plasma Phys. Control Fusion} \textbf{\bibinfo{volume}{40}},
  \bibinfo{pages}{879} (\bibinfo{year}{1998}).

\bibitem[{\citenamefont{Waelbroeck and Hazeltine}(1988)}]{Waelbroeck88}
\bibinfo{author}{\bibfnamefont{F.~L.} \bibnamefont{Waelbroeck}}
  \bibnamefont{and} \bibinfo{author}{\bibfnamefont{R.~D.}
  \bibnamefont{Hazeltine}}, \bibinfo{journal}{Phys. Fluids}
  \textbf{\bibinfo{volume}{31}}, \bibinfo{pages}{1217} (\bibinfo{year}{1988}).

\end{thebibliography}
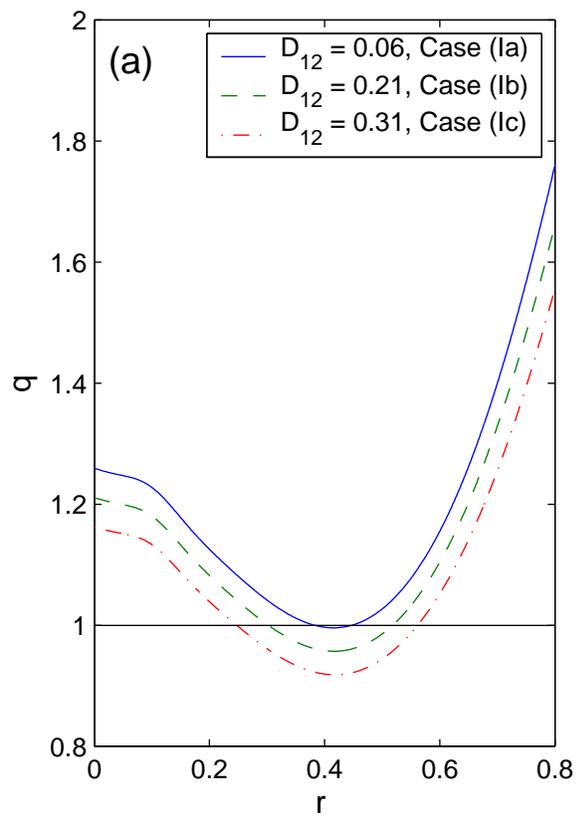
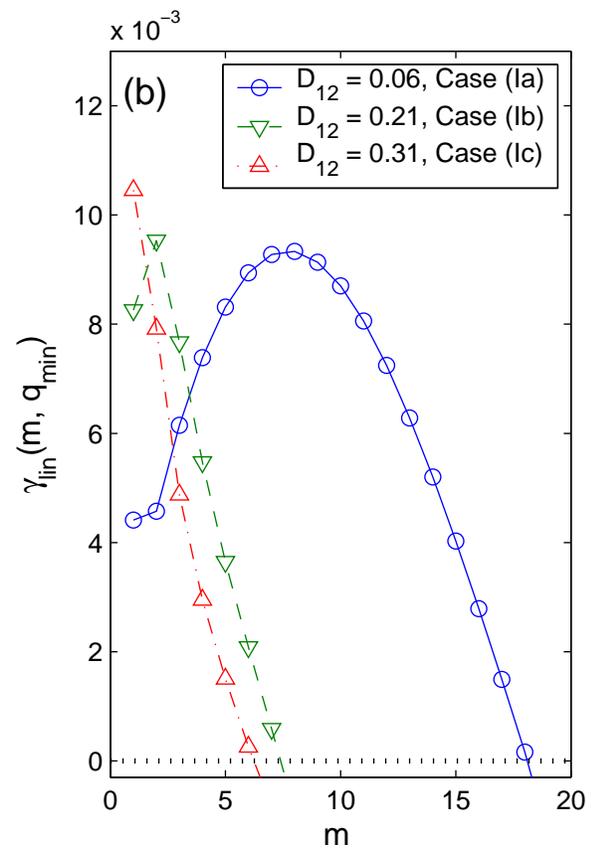
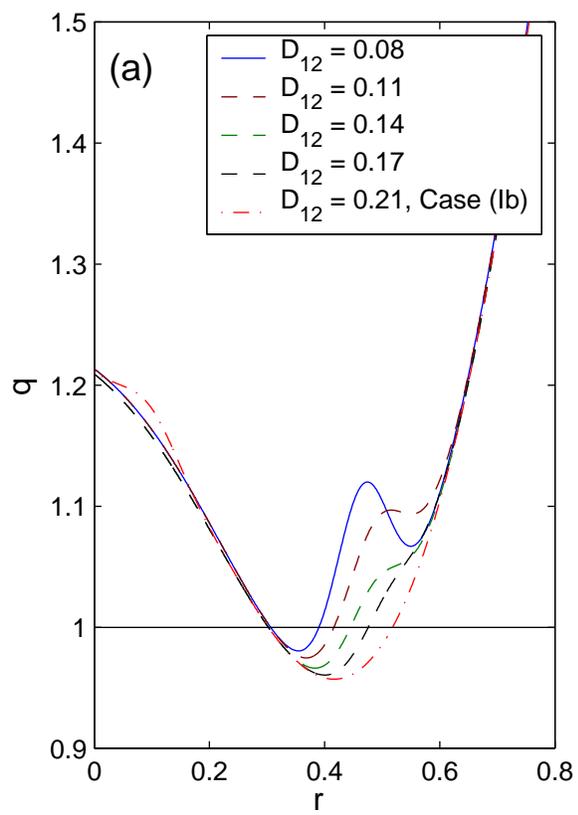
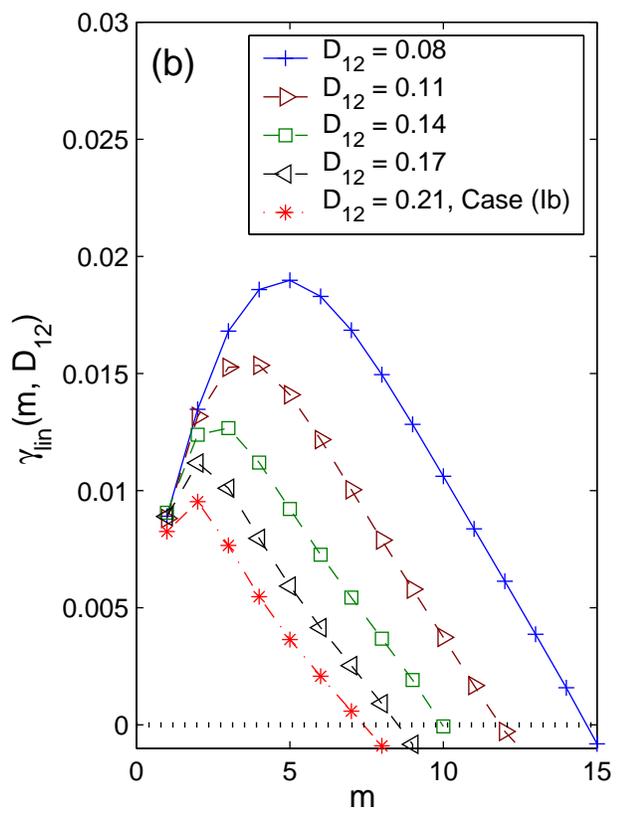
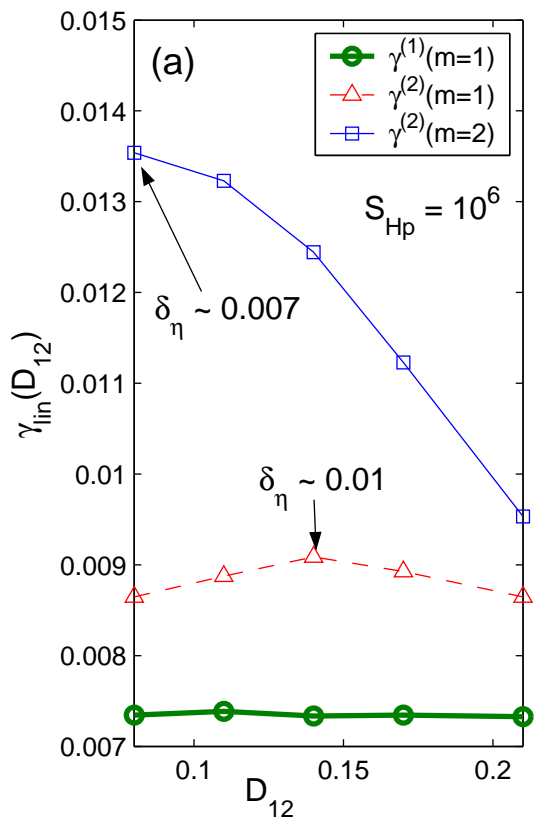
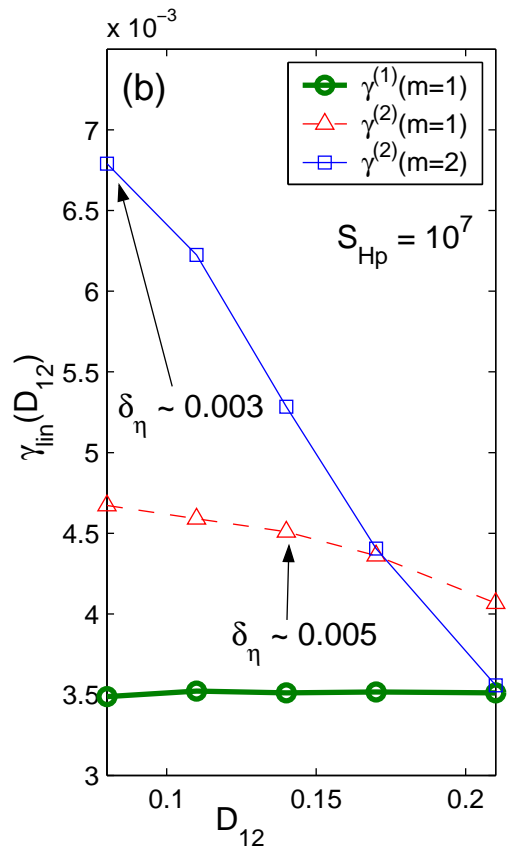
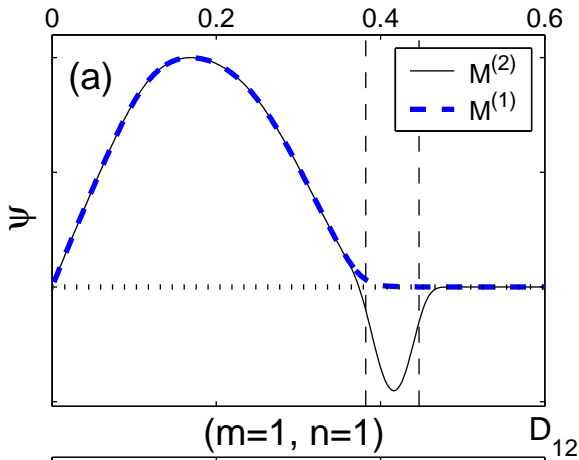
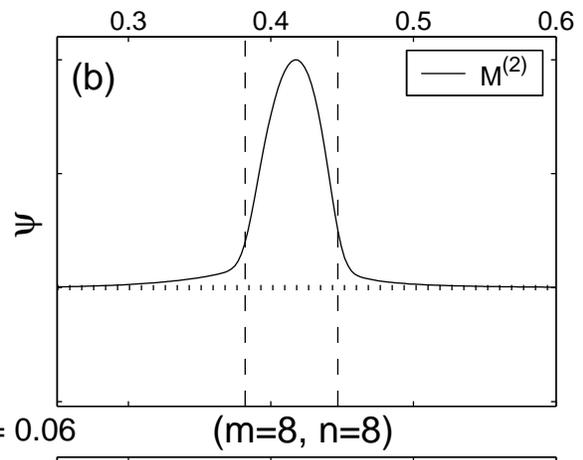
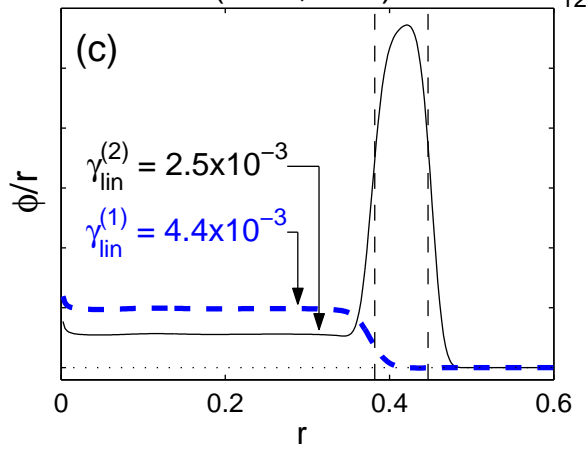
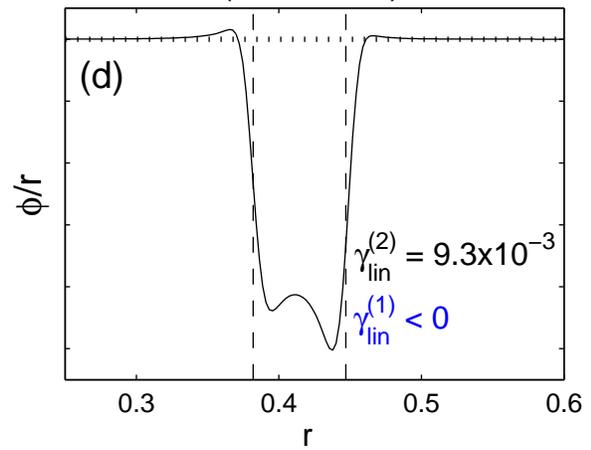
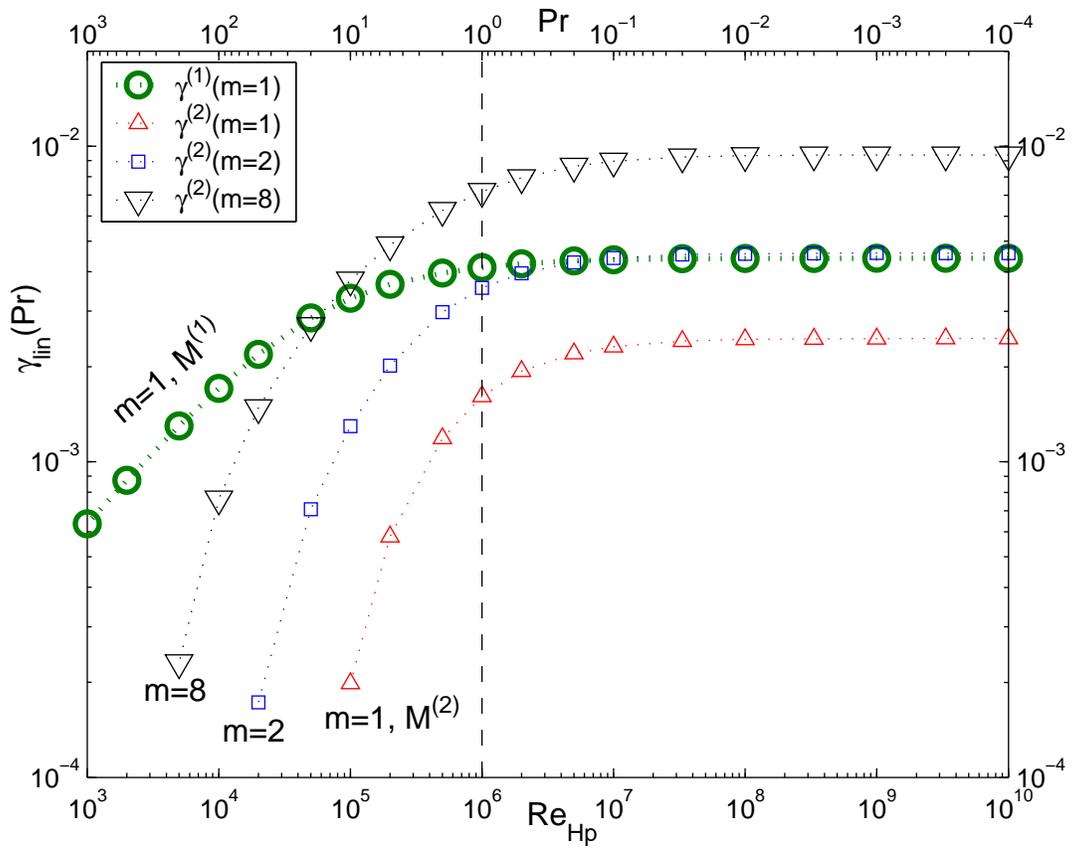
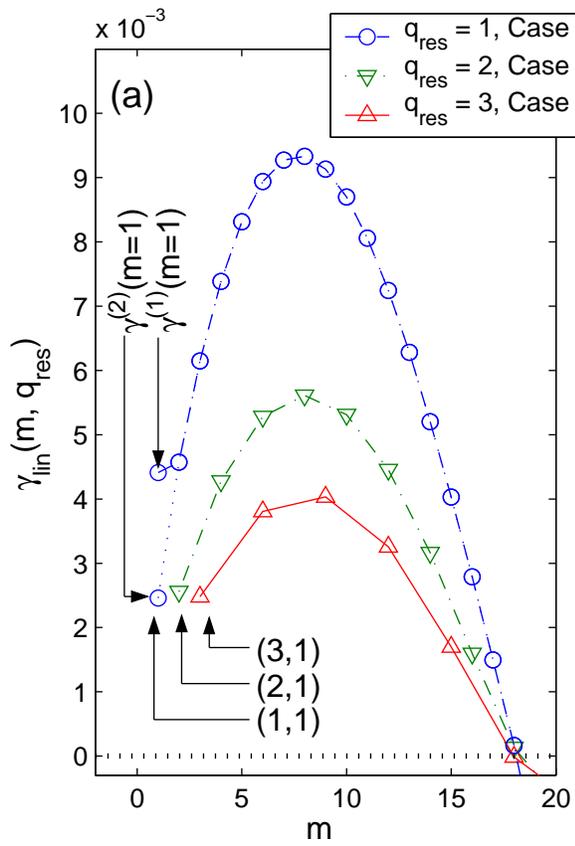
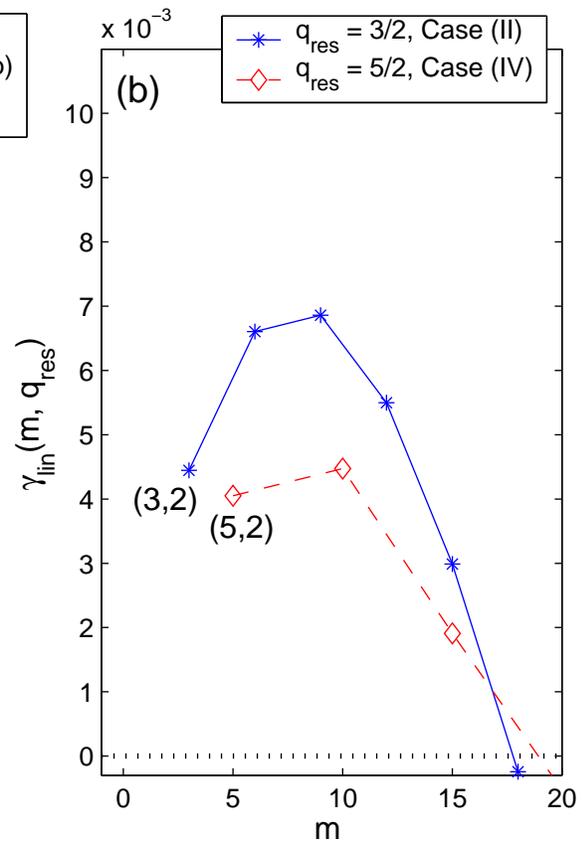
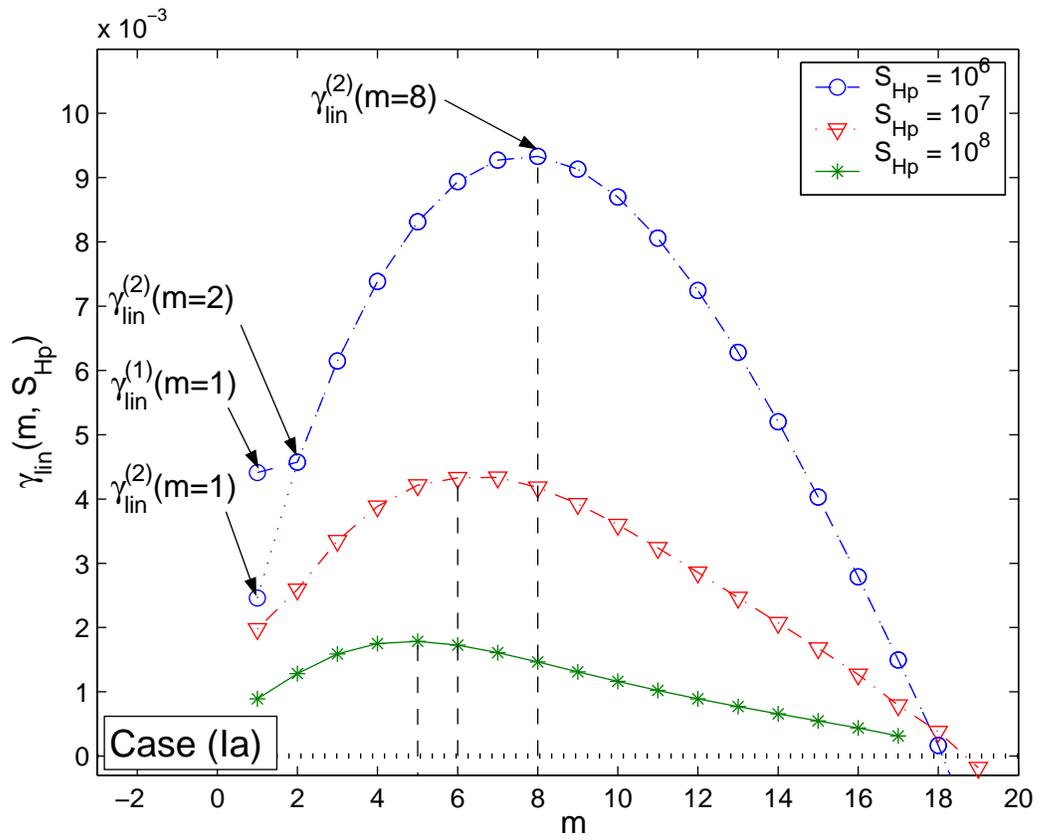
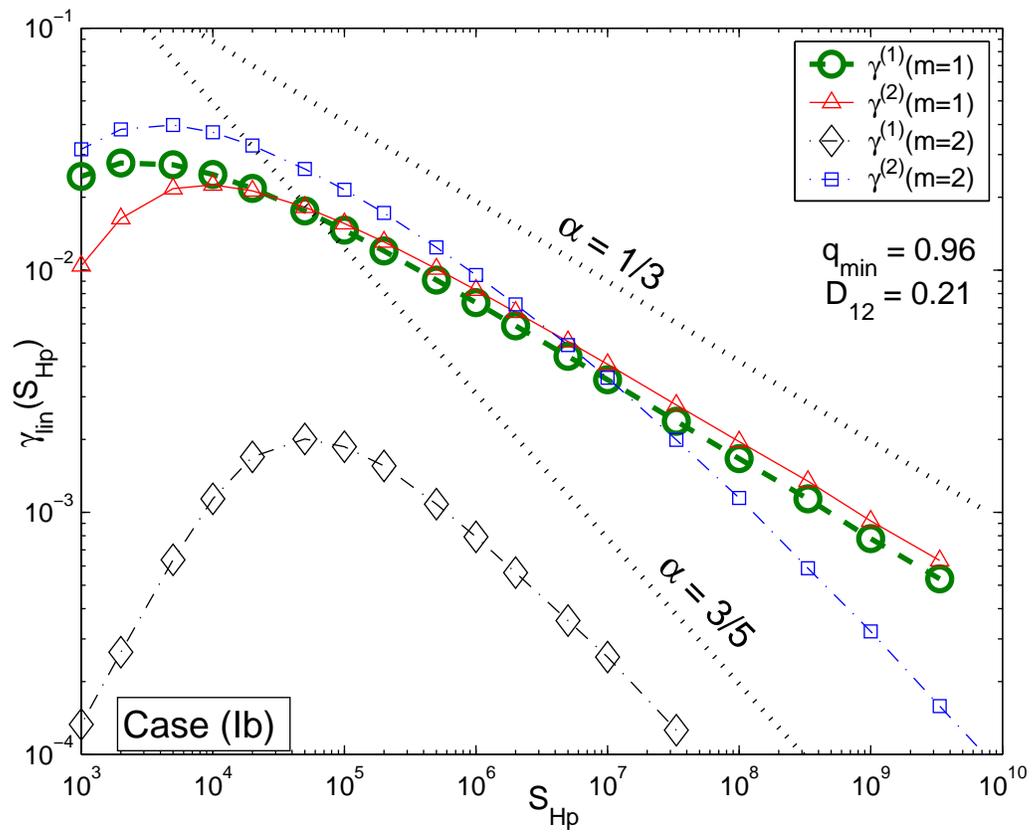
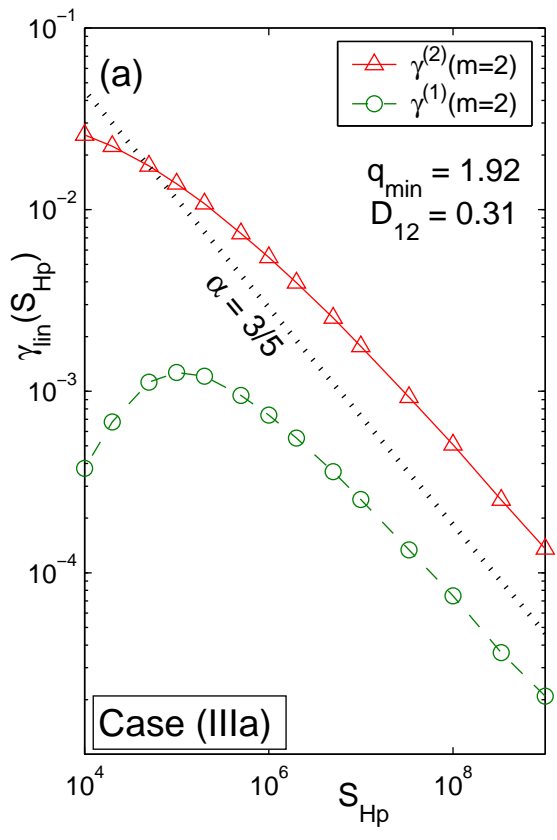
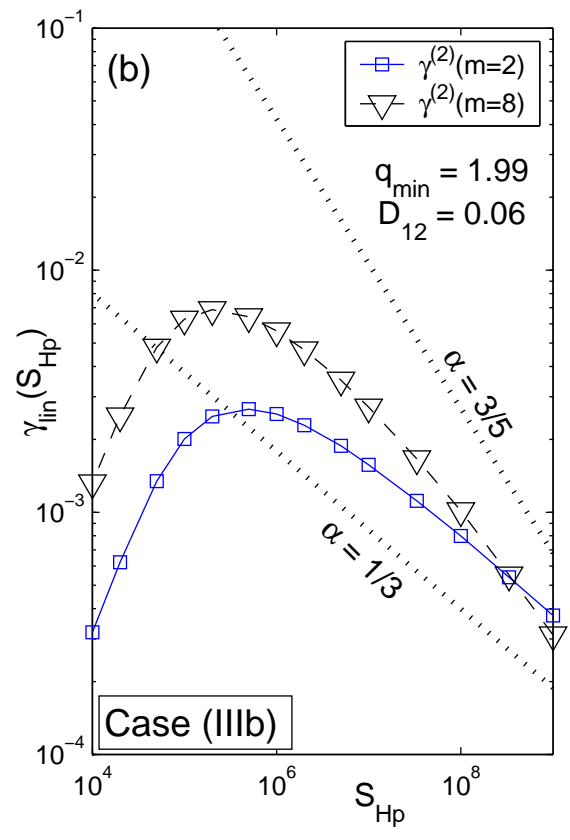
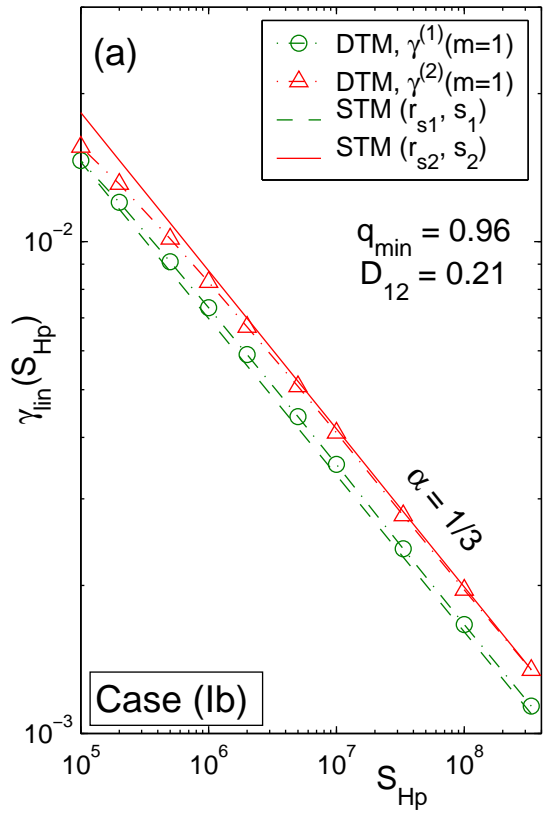
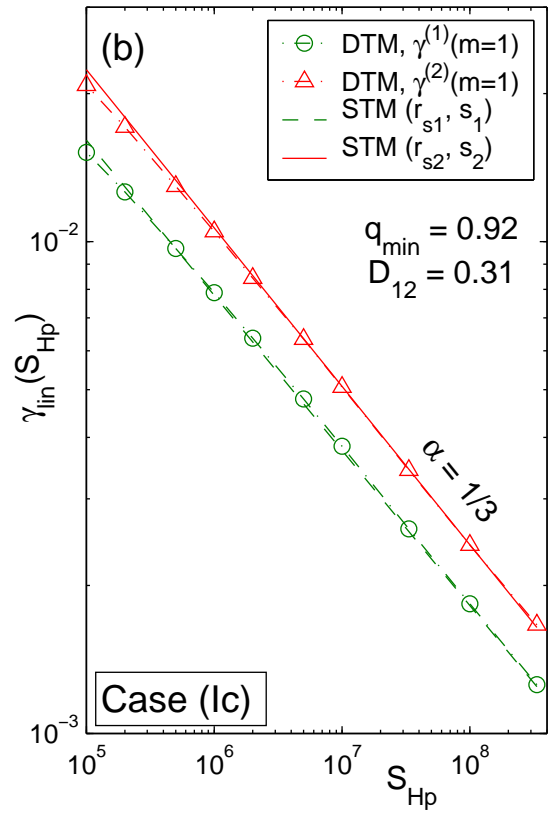
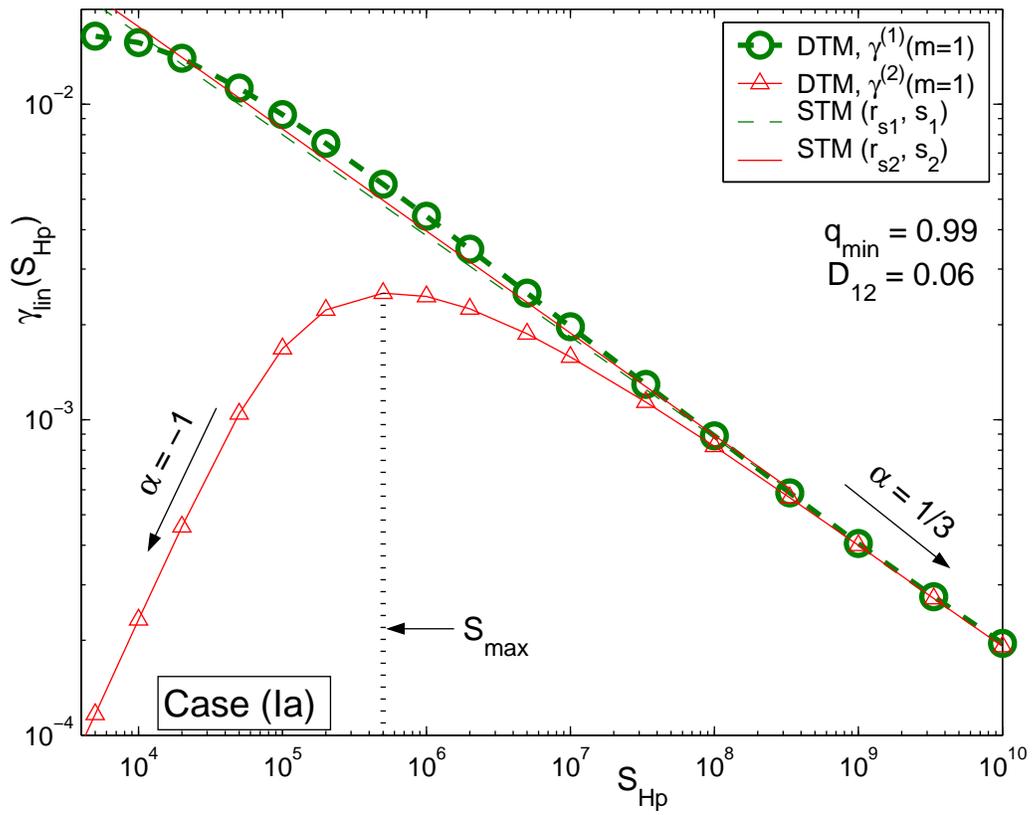
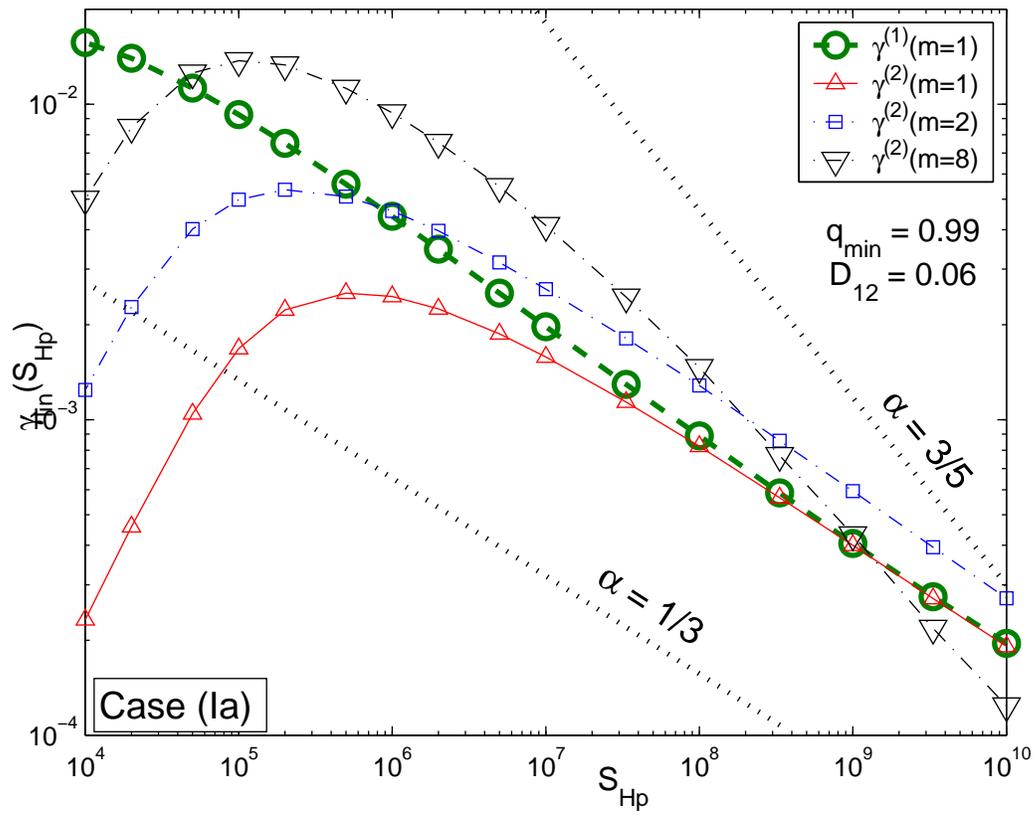
\end{document}